\documentclass[useAMS,usenatbib]{mn2e}
\usepackage{amssymb,amsmath,epsfig,times,natbib,longtable}

\long\def\symbolfootnote[#1]#2{\begingroup%
\def\thefootnote{\fnsymbol{footnote}}\footnote[#1]{#2}\endgroup} 

\title[Revealing AGN Variability]{Revealing the X-ray Variability of AGN with Principal Component Analysis}

\author[M. L. Parker et al.]{M. L. Parker,$^1$\thanks{Email: mlparker@ast.cam.ac.uk}
  A. C. Fabian,$^{1}$
  G. Matt,$^{2}$
  K. I. I. Koljonen,$^{3}$
  E. Kara,$^{1}$
  W. Alston,$^{1}$ \newauthor
  D. J. Walton,$^{4}$
  A. Marinucci,$^{2}$
  L. Brenneman$^{5}$
  and G. Risaliti$^{5,6}$\\
  $^{1}$Institute of Astronomy, Madingley Road, Cambridge, CB3 0HA\\
  $^{2}$Dipartimento di Matematica e Fisica, Universit\`{a} degli Studi Roma Tre, via della Vasca Navale 84, 00146 Roma, Italy\\
  $^{3}$Aalto University Mets\"ahovi Radio Observatory, Mets\"ahovintie 114, FIN-02540 Kylm\"al\"a, Finland\\  
  $^{4}$New York University Abu Dhabi, PO Box 129188, Abu Dhabi, UAE \\
  $^{5}$California Institute of Technology, 1200 East California Boulevard, Pasadena, CA 91125, USA\\                                                                                                 
  $^{6}$Harvard-Smithsonian Center for Astrophysics, 60 Garden St, Cambridge, MA 02138, USA\\      
  $^{7}$INAF--Osservatorio Astrofisico di Arcetri, Largo Enrico Fermi 5, 50125 Firenze, Italy\\ 
}
\date{}

\begin{document}

\maketitle

\begin{abstract}
We analyse a sample of 26 active galactic nuclei with deep \emph{XMM-Newton} observations, using principal component analysis (PCA) to find model independent spectra of the different variable components.  In total, we identify at least 12 qualitatively different patterns of spectral variability, involving several different mechanisms, including five sources which show evidence of variable relativistic reflection (MCG--6-30-15, NGC~4051, 1H~0707-495, NGC~3516 and Mrk~766) and three which show evidence of varying partial covering neutral absorption (NGC~4395, NGC~1365, and NGC~4151). In over half of the sources studied, the variability is dominated by changes in a power law continuum, both in terms of changes in flux and power law index, which could be produced by propagating fluctuations within the corona.
Simulations are used to find unique predictions for different physical models, and we then attempt to qualitatively match the results from the simulations to the behaviour observed in the real data. We are able to explain a large proportion of the variability in these sources using simple models of spectral variability, but more complex models may be needed for the remainder. We have begun the process of building up a library of different principal components, so that spectral variability in AGN can quickly be matched to physical processes.
We show that PCA can be an extremely powerful tool for distinguishing different patterns of variability in AGN, and that it can be used effectively on the large amounts of high-quality archival data available from the current generation of X-ray telescopes. 
\end{abstract}

\begin{keywords}
Galaxies: active -- Galaxies: Seyfert -- Galaxies: accretion
\end{keywords}

\section{Introduction}
Active galactic nuclei (AGN) can be extremely variable in the X-ray band on time scales as short as the light travel time over a few gravitational radii ($R_\textrm{G}$), which can be from minutes to days, depending on the mass of the black hole (BH). Examination of the X-ray spectra of AGN reveals a large variety of different spectral shapes, produced by various different processes, most notably absorption by intervening material \citep[see review by][]{Turner09Rev} and  reflection from the accretion disk and surrounding material \citep[see reviews by][]{Fabian10Rev,Reynolds13Rev}. AGN spectra can be very complex, with multiple different models providing acceptable fits to the same dataset, meaning that spectral fitting of integrated datasets alone cannot sufficiently distinguish between alternative physical models.
By investigating the properties of variability in these sources, we can hope to identify the physical processes driving AGN, probing them at different distances from the event horizon by looking at different time scales.

Principal component analysis (PCA) is a method of decomposing a dataset into a set of orthogonal eigenvectors, or principal components (PCs), which describe the variability of the data as efficiently as possible \citep[e.g.][]{Kendall75,Malzac06}. In practise, when applied to a set of spectra, this produces a set of variable spectral components which describe the variability of the source spectrum. If the spectrum is made up of a linear sum of variable, uncorrelated and spectrally distinct physical components then PCA will, with sufficient data quality, return an exact description of the physical components. 
The advantage of this method is that it produces detailed spectra of each variable component, in a model independent way. Calculating the RMS spectrum \citep[e.g.][]{Edelson02} can show the total variability as a function of energy, but cannot be used to determine how many variable components contribute to the variability or to isolate contributions from different mechanisms. Components that are only weakly variable, such as variations in absorption or reflection, will usually be drowned out by variations in the primary continuum. 
Detailed spectral modelling can be used to overcome this limitation, by carefully fitting the data for different intervals and identifying the origin of the variability. However, this is by definition not model independent, meaning that very different conclusions can be drawn from the same dataset \citep[for example, see the discussion on reflection/absorption models in MCG--6-30-15 by][]{Marinucci14}.
PCA combines the advantages of both of these methods, and can be used to calculate model independent spectra of multiple variable spectral components. This technique has many applications both within astronomy and in other fields \citep{Kendall75}, and has been used as a powerful tool for the analysis of X-ray binary variability \citep{Malzac06,Koljonen13}.

Early attempts at using PCA to understand spectral variability in AGN were hampered by a lack of high quality data. PCA has been used in a minor role for examining X-ray spectral variability in AGN for some time \citep[e.g.][]{Vaughan04}, but frequently at low spectral resolution and with only one component confidently identified. The use of singular value decomposition \citep[SVD,][]{Press86} allows the full spectral resolution of the instrument to be retained, producing detailed component spectra \citep{Miller07}. More recent work \citep{Parker14a,Parker14b} has demonstrated that PCA can return multiple components from a sufficiently large dataset, effectively isolating different spectral components.
There is now more than a decade of archival \emph{XMM-Newton} EPIC-pn \citep{Struder01} data which can be used to examine the variability of AGN on long time scales and at high spectral resolution. In this work we present a systematic analysis of 207 observations of 26 bright, variable AGN, using PCA to reveal hidden patterns of variability and to relate these patterns to the physical processes in AGN. The paper is organised as follows:
\begin{itemize}
\item In \S~\ref{datareduction} we describe the data used in this analysis and how it was processed, along with details of the analysis itself. We include a demonstration of the method with a simple toy model, showing the potential power of PCA as an analytic tool.
\item In \S~\ref{simsection} we give details of our method of simulating PCA spectra, and present the results of our simulations for different physical models of AGN variability. These different spectra then represent different predictions for the different spectral models, which we can use to understand the results from real data.
\item We present the results of the analysis in \S~\ref{resultssection}, describing and showing the PCs found for each source, along with some background on each object. We also give some basic interpretation of each result, attempting to match the PC spectra to those found using simulations.
\item Finally, in \S~\ref{discussion} and \S~\ref{conclusions}, we discuss out main results and summarise our conclusions.
\end{itemize}

\section{Observations, Data Reduction and Analysis Method}
\label{datareduction}
We restrict this analysis to \emph{XMM-Newton} EPIC-pn data only. The method used is also applicable to other instruments, but that is beyond the scope of this paper. We select only sources with more than one orbit of exposure time.
We used Science Analysis Software (SAS) version 13.0.0 for all data reduction. The data are filtered for background flares, and we use the epproc SAS task to reduce the data. We use 40 arcsecond circular regions for both source and background spectra for all sources, selecting the background region to avoid contaminating sources. Representative spectra for each source can be found in Appendix~\ref{appendix_spectra}.

The list of observation IDs used in this analysis for every source is shown in Table~\ref{obstab} (full version available online). For three sources (NGC~1365, MCG--6-30-15 and Ark~120) we have made use of data from joint \emph{XMM-Newton} and \emph{NuSTAR} \citep{Harrison13} observing campaigns \citep{Risaliti13,Walton14,Marinucci14,Matt14}. All other data is publicly available, and was downloaded from the \emph{XMM-Newton} Science Archive (XSA). To give some idea of the nature of the variability in each source and how it changes between observations we show count-count plots in Appendix~\ref{appendix_spectra}. In the majority of cases, these plots are approximately linear with a large scatter, however some sources (e.g. NGC~4051, NGC~3516) show a downturn at low counts as discussed in \citet{Taylor03}, and in some cases (e.g. RE~J1034+396) the soft and hard bands seem to be independent.

\begin{table*}
\centering
\begin{tabular}{l c c c c c c}
\hline
Source & Observation ID & Date & Duration& 0.5--2~keV rate  & 2--10~keV rate& Ratio \\
& & & (s) & (s$^{-1}$) & (s$^{-1}$) &2--10/0.5--2~keV\\
\hline
\\
MCG--6-30-15 & 	0111570101 & 2000-07-11 & 46453 & 7.67 & 2.90 & 0.38\\
&				0111570201 & 2000-07-11 & 66197 & 10.86 & 4.21 & 0.39\\

&				0029740101 & 2001-07-31  & 89432 & 15.05 & 4.82 & 0.32\\
&				0029740701 & 2001-08-01 & 129367 & 16.37 & 5.31 & 0.32\\
&				0029740801 & 2001-08-05 & 130487 & 14.98 & 4.72 & 0.32\\

&				0693781201 & 2013-01-31 & 134214 & 20.05 & 6.24 & 0.31\\
&				0693781301 & 2013-02-02 & 134214 & 11.60 & 3.86 & 0.33\\
&				0693781401 & 2013-02-03 & 48918 & 8.09 & 3.21 & 0.40\\
\\
NGC~4051 &		0157560101 & 2002-11-22 & 51866 & 2.82 & 0.61 & 0.22\\

&				0606320101 & 2009-05-03 & 45717 & 5.56 & 1.82 & 0.33\\
&				0606320201 & 2009-05-05 & 45645 & 9.37 & 2.34 & 0.25\\
&				0606320301 & 2009-05-09 & 45584 & 11.40 & 2.40 & 0.21\\
&				0606321401 & 2009-05-11 & 45447 & 8.47 & 1.69 & 0.20\\
&				0606321501 & 2009-05-19 & 41843 & 8.56 & 2.09 & 0.24\\
&				0606321601 & 2009-05-21 & 41936 & 17.58 & 3.21 & 0.18\\
&				0606321701 & 2009-05-27 & 44919 & 3.51 & 1.30 & 0.37\\
&				0606321801 & 2009-05-29 & 43726 & 4.72 & 1.81 & 0.38\\
&				0606321901 & 2009-06-02 & 44891 & 2.15 & 0.51 & 0.24\\
&				0606322001 & 2009-06-04 & 39756 & 5.25 & 1.79 & 0.34\\
&				0606322101 & 2009-06-08 & 43545 & 1.69 & 0.64 & 0.38\\
&				0606322201 & 2009-06-10 & 44453 & 4.43 & 1.46 & 0.33\\
&				0606322301 & 2009-06-16 & 42717 & 6.19 & 1.37 & 0.22\\

\hline
\end{tabular}
\caption{List of observations used in this paper. Sources are ordered by their first appearance in the text. We show 0.5--2 and 2--10~keV count rates for each observation, and the ratio between the two, so that the amplitude of spectral variability can be estimated. Note that the on-source exposure time will be smaller than the total duration, and the actual usable time used depends on the size of the intervals we use to extract spectra. Full table is available online.}
\label{obstab}
\end{table*}

In general, we follow the methods discussed in \citet{Parker14a}, hereafter P14a. For each source, we calculate the fractional deviations from the mean for a set of spectra (see example spectra for NGC~4051 in Appendix~\ref{appendix_spectra}), extracted from 10~ks intervals (unless otherwise specified). These spectra are arranged into an $n\times m$ matrix $M$, where $n$ and $m$ are the number of energy bins and spectra, respectively. We then use singular value decomposition  \citep[SVD,][]{Press86} to find a set of principal components (or eigenvectors) which describe the variability of the spectrum as efficiently as possible. SVD factorises the matrix $M$, such that $M=UAV^*$, where $U$ is an $n\times n$ matrix, $V$ is an $m \times m$ matrix, and $A$ is an $n\times m$ diagonal matrix. The matrices $U$ and $V$ then each describe a set of orthogonal eigenvectors to the matrices $MM^*$ and $M^*M$, respectively. These eigenvectors represent the spectral shape of the variable components. The corresponding eigenvalues are given by the diagonal values of $A$, and are equal to the square of the variability in each component (in arbitrary units). The fractional variability in each component can then be found by dividing the square roots of the eigenvectors by the sum of all the square roots.

The resulting components show the strength of the correlation between energy bins, so a flat positive (or negative, the sign of the y axis is arbitrary) component shows that all bins vary equally, whereas a component that is positive at low energies and negative at high energies represents a pivoting effect. This is complicated by the requirement that the eigenvectors are orthogonal, i.e. their dot-product must be zero. We examine the effect of this constraint on simulated PCs in \S~\ref{simsection}. 
We note that the method of preparing the spectra is such that constant multiplicative components will have no effect on the PCA results, as they will not affect the fractional residuals. However, a constant spectral component that changes with energy will suppress the spectral variability of the variable components. This has important implications for distinguishing between absorption and reflection in AGN variability.

\begin{figure}
\centering
\includegraphics[width=8cm]{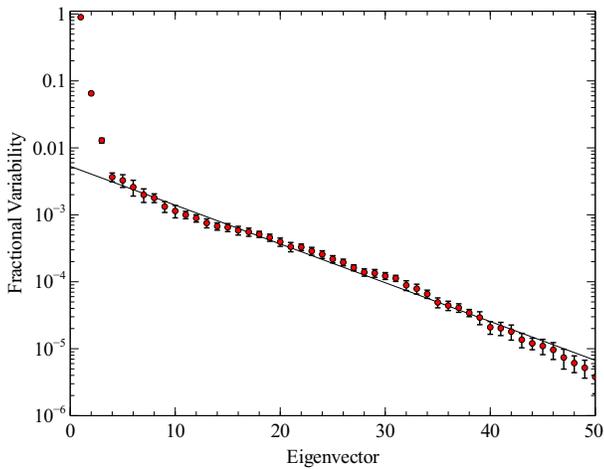}
\caption{LEV diagram for Ark~564. This shows the fractional variability in each eigenvector obtained from the PCA of this source. The black line shows the best-fit geometric progression, fit from component 4 to 50. The remaining three components are found to be highly significant, as they deviate from this line by many standard deviations.}
\label{ark564eigenvectors}
\end{figure}

The significance of the components produced is determined using the log-eigenvalue (LEV) diagram, an example of which (for Ark~564) is shown in Fig.~\ref{ark564eigenvectors}. This shows the fraction of the total variability which can be assigned to each PC, so for Ark~564, $\sim90$ per cent of the variability is in the first PC and so on. The components due to noise produced by the PCA are predicted to decay geometrically \citep[see e.g.][]{Jolliffe02,Koljonen13}, so deviation from a geometric progression can be used as a test of the significance of a component. In this case, three components deviate from the best-fit geometric progression, and are highly significant. For the sake of brevity, we only show an example LEV diagram, rather than one for every source.
The strongest statement about the significance of the components we investigate comes simply from the strong correlation between points in adjacent bins. Any coherent components produced are extremely unlikely to be due to random noise, which is independent between bins.

\begin{figure*}
\centering
\includegraphics[width=17cm]{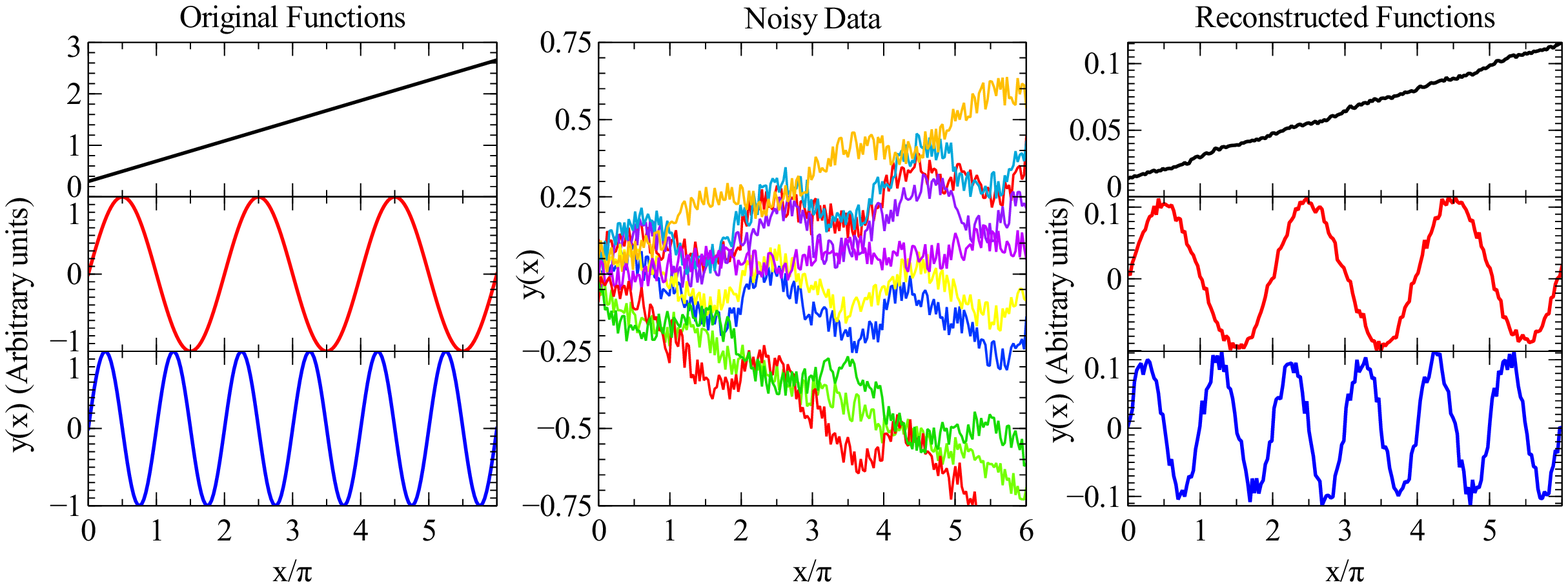}
\caption{Left: The three functions we add together, along with noise, as a simple test of the PCA method. Middle: Sample of 10 `spectra' created by adding together the three functions on the left in random amounts with additional noise. Right: The PCs returned by the analysis. Note that the y axis is arbitrary and different between the left and right panels. Error bars are not plotted for clarity.}
\label{reconstruction}
\end{figure*}

\begin{figure}
\centering
\includegraphics[width=8cm]{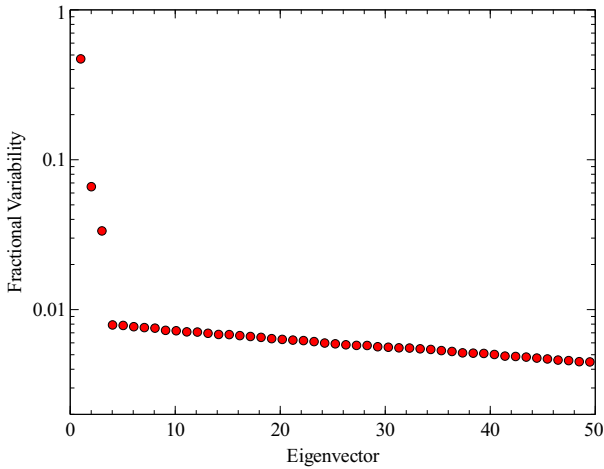}
\caption{LEV diagram for the test case shown in Fig.~\ref{reconstruction}. In this test, three input functions are summed, and noise is added. The LEV diagram shows that three returned components are statistically significant, the remainder can be attributed to noise. Error bars are not shown, but are smaller than the points for the three significant eigenvectors.}
\label{testeigenvectors}
\end{figure}

A simple test case is shown in Fig.~\ref{reconstruction}. For this example, we add together three functions: $y_1(x)=0.3+x/8$; $y_2(x)=\sin(x)$ and $y_3(x)=\sin(2x)$, along with random noise. We create fifty `spectra', of the form $y(x_i)=0.6a_1y_1(x_i)+0.2a_2y_2(x_i)+0.1a_3y_3(x_i)+0.1a_4$, where $a_j$ are random values, evenly distributed between $\pm0.5$ and $x_i$ are the 200 values between 0 and $6\pi$ over which the functions are calculated. In the left panel we show the three input spectra, minus noise, in the middle panel we show a sample of the generated functions and in the right panel we show the functions recovered using our PCA code. The LEV diagram for this test is shown in Fig.~\ref{testeigenvectors}, and clearly shows that three components are significant, with 47 per cent of the variability in the first component, 7 per cent in the second, and 3 per cent in the third. All the remainder is attributable to noise. These values are functions of the amplitude and variability of each component, and the signal to noise ratio of the data.

P14a calculated extremal spectra and used comparisons to spectral fitting to find physical interpretations for the components produced in that analysis. However, this is inefficient for large samples of objects and could potentially compromise the model independence of the results. In this work, we create simulated spectra (see \S~\ref{simsection}) based on physical models that are allowed to vary within given parameter ranges, then use PCA to find the component spectra for each model. This produces a predicted set of PC spectra for each model, which we can then match to the PC spectra found from the data for each source.

\section{Simulations}
\label{simsection}
In this section we use simulations to predict the PCA spectra produced by different models of AGN variability. This technique for analysing PCA spectra was used by \citet{Koljonen13}, and the method we use here was introduced in \citet{Parker14b}, hereafter P14b, where it is used to demonstrate the differing predictions for intrinsic source variability and absorption variability in NGC~1365.

\subsection{Method}
In general, we follow the method outlined in P14b and simulate spectra using the xspec command `fakeit', then analyse the results using the same method as we use for the real data. The parameters of interest are selected randomly between extreme values for each spectrum.
For simplicity, we simulate 10~ks \emph{EPIC-pn} spectra, to be as similar as possible to those found from real data. We do not match the model flux to the data, instead exaggerating the model flux so that the features are more prominent. This is equivalent to simulating longer exposures at lower flux for the simple models discussed here, but less time consuming to calculate.
We do not attempt to exactly match the PCs produced by the data, instead looking to produce general predictions for different variability mechanisms. 

All of the components produced by these simulations are equally valid with the y axis inverted as they represent deviations from the mean, rather than the minimum, and will therefore sometimes be positive and sometimes negative. In general, we attempt to arrange the components in the manner which makes the most intuitive sense.

We will initially consider the PCs returned from a variable powerlaw, and then investigate the effect of additional spectral components. In each case, we will first look at the effect of including a constant component and varying the powerlaw, then allowing the new component to vary.  Finally, we will examine a select few examples  with more than one additional variable component.

Where the components returned from the simulation correspond directly to one of the model components, we label the figure with the relevant symbols: $N_\textrm{pl}$, $N_\textrm{bb}$, $N_\textrm{ref}$, $N_\textrm{H}$ and $f_\textrm{cov}$ correspond to power law, black body and reflection normalizations, column density and covering fraction respectively.

\subsection{Single and multiple power laws}
\label{section_plsims}

\begin{figure}
\centering
\includegraphics[width=8cm]{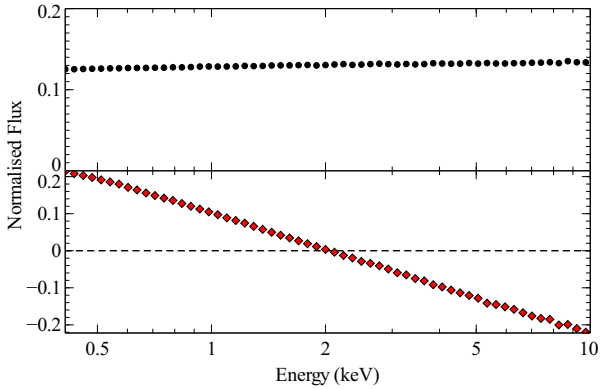}
\caption{The first (top, black) and second (bottom, red) principal components obtained by using PCA on simulated spectra of a power law which varies in normalisation and photon index. The first component corresponds to changes in the normalization, and the second to pivoting of the power law.}
\label{powerlawsims}
\end{figure}

As a baseline model, we establish the components expected from variability of a power law continuum. Fig.~\ref{powerlawsims} shows the two components obtained from a simulation of a simple variable powerlaw, with no other spectral components. The photon index is allowed to vary between 1.9 and 2.1 randomly, and the normalisation is allowed to change by a factor of two. The resultant components are completely straight, showing no features of any kind, although there is a slight increase with energy in the primary component, due to a correlation between flux and photon index in the model. 

For a good comparison with real data, we refer the reader to 3C~273 in \S\ref{section_otherthings}. This source is dominated by a powerlaw from a relativistic jet, and the first two components found from the data are an excellent match to the predictions for a varying powerlaw shown in Fig.~\ref{powerlawsims}.

\begin{figure}
\centering
\includegraphics[width=8cm]{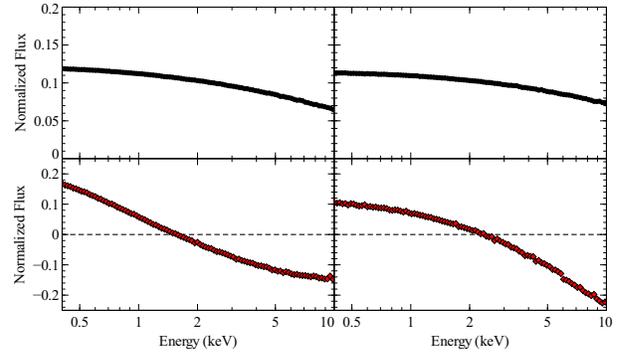}
\caption{Left: The two significant PCs from a simulation of a power law varying in normalisation and photon index, in the presence of a constant hard powerlaw. This spectral component suppresses the variability of the PCs at high energies. Right: The two PCs from a simulation where the spectral pivoting arises from two power laws, one hard, one soft, varying in normalisation but not in index. These components are very similar to those produced by a pivoting power law, but with additional curvature.}
\label{sims_twopls}
\end{figure}

We next investigate the effect of adding a second power law to the spectrum. Additional continuum components  such as this are hard to distinguish spectrally, but have been suggested by some studies \citep[e.g.][]{Grupe08,Noda13} and are a natural consequence of multi-zone Comptonization models. We therefore include a weaker second power law, with a harder photon index $\Gamma=1$ and a normalisation of 0.1 times that of the primary powerlaw. When we keep this second continuum component constant and vary the primary power law as before, the effect is simply to lower the fractional variability of the primary component with increasing energy. This is shown in the left two panels of Fig.~\ref{sims_twopls}.

We have so far treated spectral pivoting as being due to changes in the photon index of a primary power law continuum. However, it is possible to generate a similar effect from the interplay between two (or more) continuum components with different photon indices changing in normalization. In the right two panels of Fig.~\ref{sims_twopls} we show the two components produced from the same two power law model, when both power law components change in normalisation but not index. The primary power law is varied in normalization between 0.5 and 2, and the secondary powerlaw between 0.08 and 0.12. As expected, the first PC is very similar in this case and that with no variability in the second power law, but there are qualitative differences in the second component. As this now corresponds to the spectral pivoting caused by changes in the flux of the second continuum component, it is enhanced at high energies, rather than damped out as in the previous case. This may be relevant to the objects in \S\ref{Ark564section}, where the second component gets steeper with energy. For the remainder of these simulations, we will only consider the case of a single, pivoting powerlaw, but the reader should bear in mind that a similar effect could be achieved with a combination of two or more such components that change in relative amplitude rather than index.

\subsection{Black body/soft excess components}
We now investigate the PCs produced from a constant or weakly variable soft excess component. For our simulations we use a black body for simplicity and brevity, but the resulting components are equivalent to those that would be produced by any other models that explain the soft excess in terms of an additional component that only contributes at low energies (e.g. Comptonization and Bremsstrahlung models).

\begin{figure*}
\centering
\includegraphics[width=12cm]{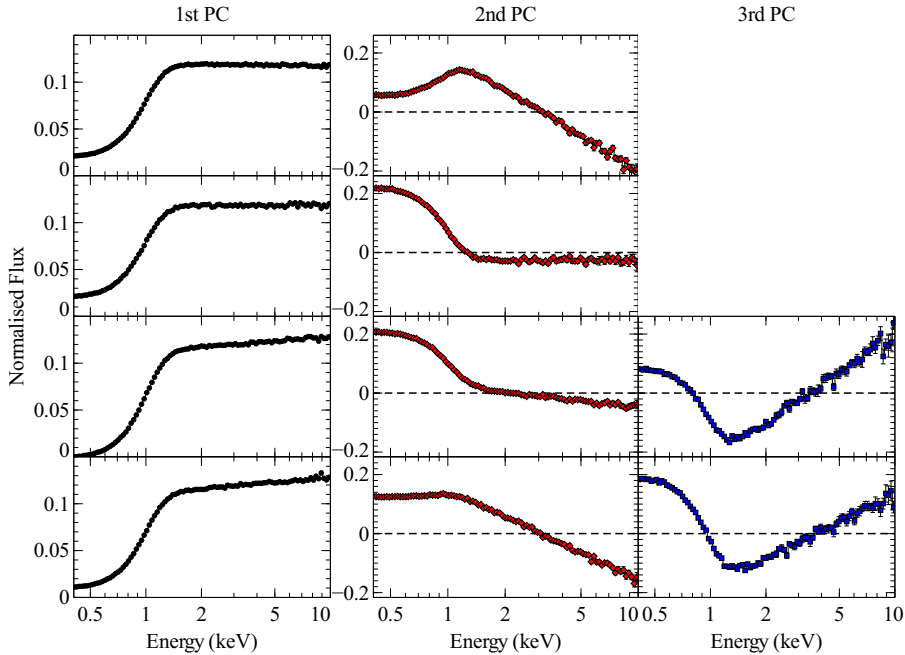}
\caption{PCs returned from simulations of power law and black body variability. The top row shows the components produced when the power law is allowed to vary in normalisation and photon index, while the black body is kept constant. The second row shows those for a fixed photon index, but where the black body flux is allowed to vary. The third and fourth rows show the components produced when all three parameters are allowed to vary, with the black body variations being stronger in the third and the power law pivoting dominating in the fourth.}
\label{bbsims}
\end{figure*}

Initially, we consider the effects of a constant black-body component on the PCs returned from a varying power law. For this simulation we include a black body with a temperature of 0.1~keV and a normalisation of 0.1, then vary a power law with a photon index of between 1.9 and 2.1, and normalisations between 0.5 and 1.0. The effect of such a constant component is to suppress the variability seen in the variable components, pushing the bins where the black-body is strongest towards zero. This can be seen in the top row of Fig.~\ref{bbsims}, where the flat PCs produced by a varying power law are pushed towards zero at low energies by the black body. 

Fixing the photon index and allowing the black body to vary (between normalisations of 0.8 and 1.2) produces a different second component, (top middle panel of Fig.~\ref{bbsims}) which has the same shape as the black body at low energies, then is negative but close to zero at high energies. The negative values are caused by the orthogonality constraint of PCA, which requires that the dot-product of any two PCs be zero. In practise, this means that if the primary component is 100 per cent positive, approximately 50 per cent of the energy bins of all subsequent PCs must be below zero.

\begin{figure}
\centering
\includegraphics[width=8cm]{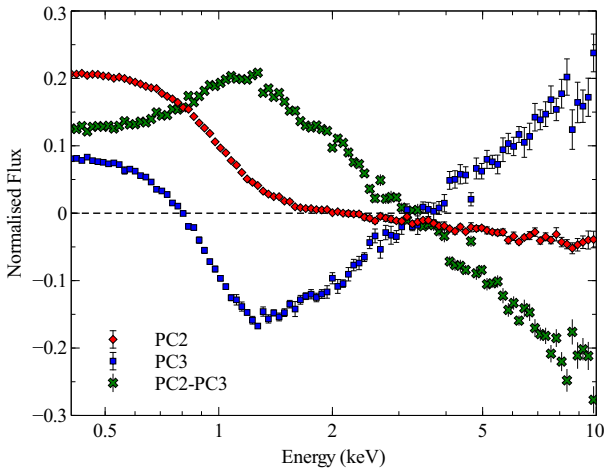}
\caption{The effect of subtracting the `correction factor' PC from the second order black body variability term for the simulations of power law and black body variability. The PC corresponding to the black body variations is shown in red diamonds, and the third PC is shown in blue squares. When one is subtracted from the other, the resulting component (green crosses) matches the pivoting PC produced when the photon index variations dominate (see the top row of Fig.~\ref{bbsims}).}
\label{bbpcs_added}
\end{figure}

A more complex component is produced if we vary the photon index of the power law as well. Firstly, we vary the index weakly (between 1.95 and 2.05), and the results of this are shown in the third row of Fig.~\ref{bbsims}. This produces a minor change in the first two components - they both show an incline at high energies, rather than being completely flat, and produces a third significant PC (right panel). This component appears to act as a correction factor to the second  component, which no longer describes all of the pivoting itself. If we double the range that the photon index varies over, we see that the second component changes significantly (bottom row). This component is most similar to the pivoting component produced when the black body is constant (top row). The third component changes only very slightly, and shows the same general structure. Again, the third component here is a correction factor, rather than a direct match to a single physical component. However, in this case the third PC is used to make the second PC appear more like the black body PC in rows 2 and 3, rather than the other way around. In Fig.~\ref{bbpcs_added} we show the effect of adding the `correction factor' component onto the second order black body component from the weakly varying powerlaw index case. This produces a component with the same spectral shape as the pivoting term (top middle panel of Fig.~\ref{bbsims}).

The third components produced by these simulations demonstrate a key weakness of this kind of analysis - if two physical components have a similar effect on the spectrum, then they will not be expressed as two separate PCs - rather there will be one PC describing the average effect, and one describing the differences between the two. In this case, both an increase in the black body flux and an increase in the photon index produce a steeper spectrum, so the second component is an average of these two effects. This was also found to be the case in the absorption simulations shown in P14b, where because the low energy spectrum of NGC~1365 was dominated by diffuse thermal emission the absorption and intrinsic variability components were very similar, leading to an averaging effect. However, the dominant driver of the spectral variability can still be identified, as shown here and in P14b.

While we do not find any sources that show such simple black body variability (which is far more likely to be visible in X-ray binaries than AGN), it is instructive to note that variations in a soft spectral component can result in a PC that shows apparent hard variability. This may be relevant to the fourth order PC in MCG--6-30-15 and similar objects (\S\ref{mcg6objects}).

\subsection{Distant Reflection}
\label{section_distref}

Distant or neutral reflection is found in many AGN \citep[e.g.][]{Ricci14}, and occurs when X-ray emission from the corona is scattered and reprocessed by cold material, far from the black hole. Because of this much larger spatial scale, the variability in this spectral component is much lower than that found in components that originate from the inner disk, or those due to intervening clouds or winds. It follows that the main effect of distant reflection will be to damp out the variability in the energy bands where it is strong, particularly the 6.4 Fe K$\alpha$ line.

\begin{figure}
\centering
\includegraphics[width=8cm]{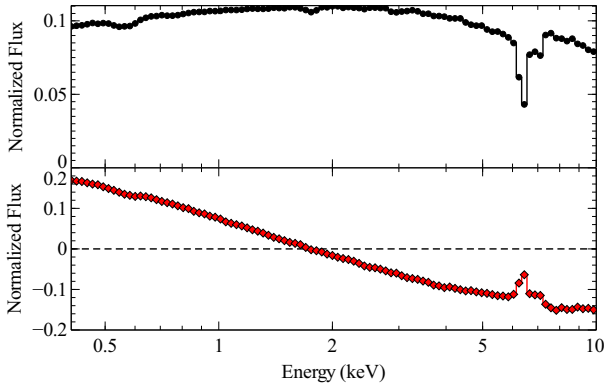}
\caption{The first (top) and second (bottom) PCs from a simulation of a power law continuum varying in normalization and spectral index in the presence of a constant neutral reflection component. The main difference caused to the shape of the components returned is the strong narrow iron line feature at 6.4~keV.}
\label{sim_distref}
\end{figure}

We show in Fig.~\ref{sim_distref} the two PCs returned from a simulation of a varying power law continuum and constant distant reflection. We model the reflection component with the \textsc{xillver} model \citep{Garcia13}, with an input power law index of 2, an iron abundance of 1, the ionization $\xi$ at the lowest allowed value of 1, and an inclination of 30 degrees. The flux of the reflection component is fixed, and is approximately equal to half the average flux of the continuum. The primary power law is then varied as before. While the resulting PCs do show slight differences in curvature over the whole energy range, these are likely to be undetectable due to noise  and the presence of other spectral components in the real data. By far the largest difference is the strong, narrow iron line visible at 6.4~keV, which suppresses the variability of the power law components, pushing them towards 0. We note that in the case where the primary power law is heavily obscured then it is probable that distant reflection will leave more signatures in the PC spectra at low energies.

\subsection{Blurred Reflection}
\label{refsimssection}
We now present the results of simulations including relativistically smeared reflection from the inner accretion disk.
P14b present the two components produced from a simulation of a varying powerlaw, and the three components found when a weakly variable (approximately 0.4 times as variable as the powerlaw) relativistic reflection component is added. We reproduce and expand upon these results here, investigating the effects of a strongly blurred and ionized reflection component.

\begin{figure}
\centering
\includegraphics[width=8cm]{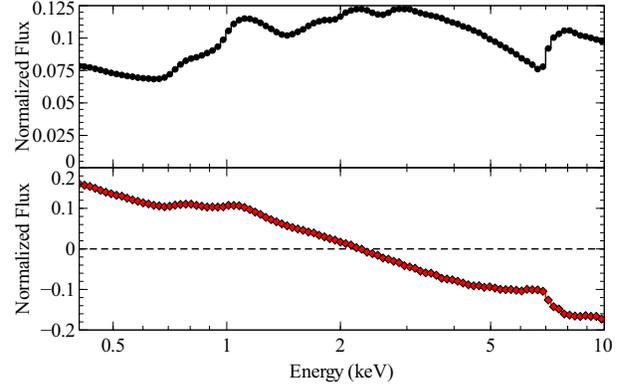}
\caption{The first two (top and bottom, respectively) PCs recovered from a simulation of a power law varying in normalization and photon index, in the presence of a constant relativistically blurred reflection component. The effect of this component is to suppress the spectral variability at the energies of the iron line and soft excess.}
\label{constreflectionsim}
\end{figure}

In Fig.~\ref{constreflectionsim} we show the two components returned from a simulation of a varying power law and constant reflection. The reflection parameters are the same as those in \S\ref{section_distref}, except the normalisation of the reflection component is increased so that the 0.3--10~keV flux is approximately equal to that of the power law. The reflection spectrum is then convolved with the \textsc{kdblur}, with the inner and outer radii set at 1.235 and 400 gravitational radii, respectively, and an emissivity index of 3. The first PC, shown in the top panel, can essentially be thought of as a flat line with the reflection component subtracted. This then represents the suppression of variability at energies where the relativistic reflection makes a substantial contribution to the total flux. 
The second component is similar, although starting from from a diagonal rather than flat line, it is likewise pushed towards zero where the reflection component dominates.

\begin{figure*}
\centering
\includegraphics[width=13cm]{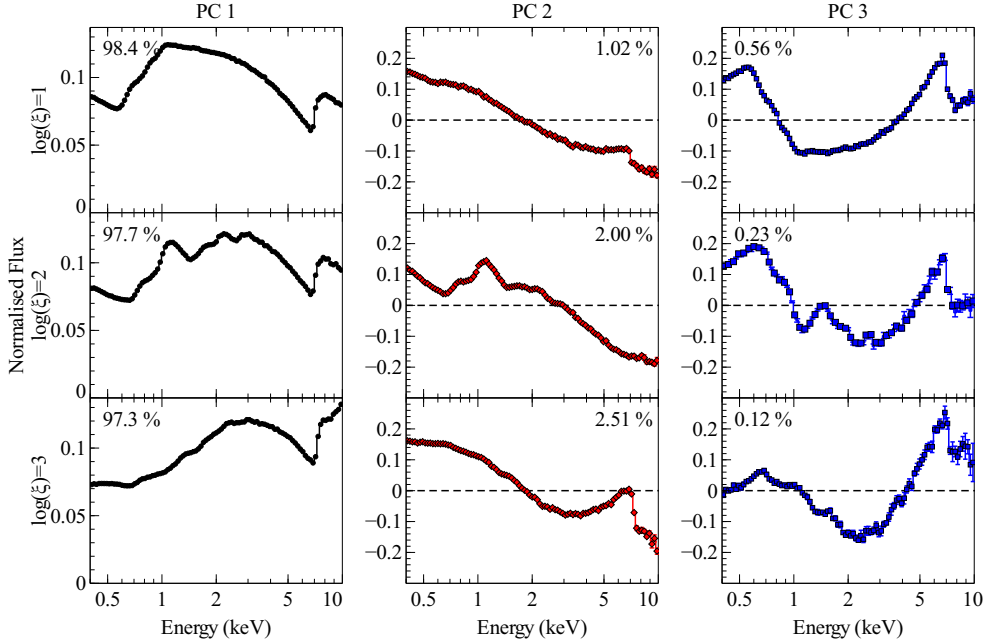}
\caption{The first (left, black), second (middle, red) and third (right, blue) principal components obtained by using PCA on simulated spectra of a power law plus relativistic reflection model, where the powerlaw flux varies by a factor of 4 and the reflected flux by a factor of 1.5. These components are shown for four different ionisation parameters, and the fractional variability attributed to each component is shown in each plot. Black dashed lines indicate zero on the y axis for the second two plots (axis labels are not shown, as the scale is arbitrary).}
\label{reflectionsims}
\end{figure*}

Fig.~\ref{reflectionsims} shows the three significant components obtained from simulations of a model with a varying power law and varying relativistic reflection, for a range of ionisation parameters. The fluxes of the power law and reflection components are kept approximately equal, but the reflection component is only allowed to vary by a factor of 2, compared to 5 for the power law. The first two PCs are equivalent to those shown in Fig.~\ref{constreflectionsim}, and are largely due to the power law variability. The additional third PC represents all of the reflection variability that cannot be adequately described by the first two PCs. This component displays the correlated soft excess and broad iron line typical of relativistic reflection. Unlike the distant reflection discussed in \S\ref{section_distref}, relativistic reflection makes a strong contribution at soft energies, and has a much broader iron line.

These simulations are particularly relevant to the sources discussed in \S\ref{mcg6objects}, 5 of which show both the low and high energy breaks in the first and second components and a higher order PC, very similar to those presented in Fig.~\ref{reflectionsims}, with a correlated soft excess and broad iron line.

\subsection{Neutral Absorption}
\label{abssims}
We are also interested in the effects of absorption variability in AGN spectra, although there are problems with using PCA in an absorption dominated variability regime. One of the key assumptions of PCA is that the dataset can be expressed as a linear sum of principal components. This assumption is reasonable when applied to additive components such as a reflection spectrum, but is not valid when we consider variable multiplicative components applied to a variable spectrum, potentially leading to spurious terms being produced. Constant absorption produces no such problem, as a constant multiplicative factor makes no difference to the fractional deviations we use to calculate the PCs. Nevertheless, as shown by P14b and as we demonstrate here, it is possible to find physically meaningful components from such an analysis. We stress that in all cases, the shape of the underlying spectrum is unimportant, provided that it is relatively constant compared to the absorption variability. Again, this is due to the PCs being calculated from the fractional residuals rather than the total spectrum.

\begin{figure}
\centering
\includegraphics[width=8cm]{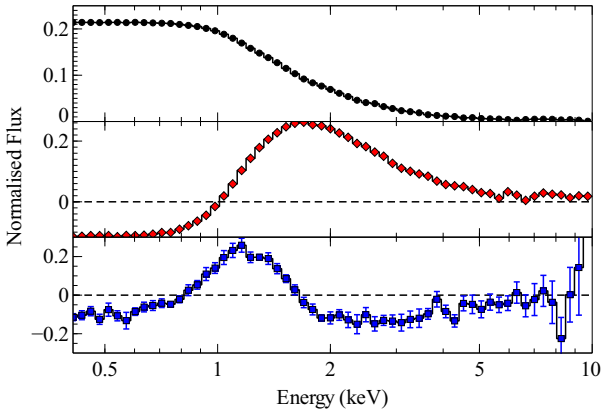}
\caption{The three principal components produced by a simulation of varying absorption. A power law is convolved with a neutral partial covering absorption model, and the column density and covering fraction of the absorber are allowed to change. The first term shown here corresponds well to the covering fraction, and the second two terms represent correction factors to this, which depend on the column density.}
\label{varyingabs1}
\end{figure}

\begin{figure}
\centering
\includegraphics[width=8cm]{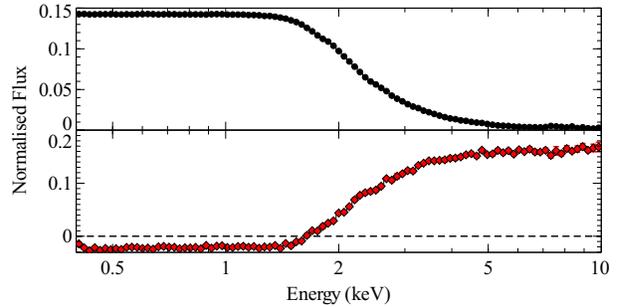}
\caption{The two principal components found from a simulation of partial covering absorption and intrinsic variability.}
\label{varyingabs2}
\end{figure}

Fig.~\ref{varyingabs1} shows the three PC spectra produced when we consider a partial covering absorption component (modelled with \textsc{zpcfabs} in Xspec) applied to a $\Gamma=2$ powerlaw. The covering fraction is allowed to vary randomly between 0 and 1, and the column density is allowed to vary by a factor of 3. The first component correlates well with the covering fraction, however the other two components returned by the analysis do not correspond directly to a single parameter and represent changes in the column density at different covering fractions. The first component matches well with the first component of NGC~4395 (\S~\ref{sec_ngc4395}), which shows strong absorption variability \citep{Nardini11}. Indeed, if we simulate a source where the variability is dominated by changes in the covering fraction of a neutral absorber, allowing some variability of the underlying powerlaw, we obtain two components which are in excellent agreement to those shown found for NGC~4395. This simulation is shown in Fig.~\ref{varyingabs2}, for a partial covering absorber that varies in covering fraction from 0.5 to 1, with $N_\textrm{H}$ fixed at $10^{22}$~cm$^-2$, applied to a power law with $\Gamma=2$ that varies by a factor of $\sim20$ per cent.

The marginally more complex simulations from P14b showed the effects of diffuse thermal emission on the PCs returned from partial covering. This addition, which characterises the PC spectra of NGC~1365, damps out the variability at low energies.

\subsection{Ionized absorption}
\begin{figure}
\centering
\includegraphics[width=8cm]{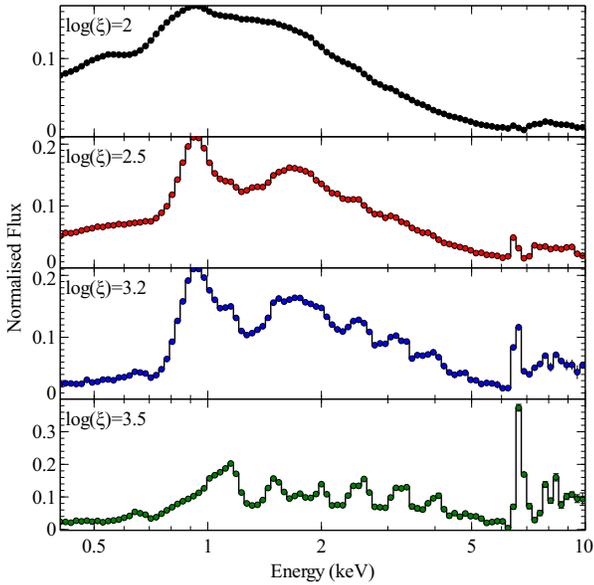}
\caption{Principal components produced by simulations of partially covering ionised absorption. The covering fraction is allowed to vary randomly between 0 and 1, with all other parameters fixed, producing a single significant component. The four panels correspond to four different ionisation parameters. This figure is reproduced from \citet{Parker14b}.}
\label{zxipcf_fig}
\end{figure}

In Fig.~\ref{zxipcf_fig} (reproduced from P14b) we show the PCs produced by variations in an ionised partial covering absorber. For this simulation we use the \textsc{zxipcf} model \citep{Reeves08} and allow the covering fraction to vary randomly between zero and one for different values of the ionisation $\xi$. This produces a single PC in all cases, the spectral shape of which depends strongly on the ionisation. We also investigate allowing the column density to vary, keeping the covering fraction fixed at 0.5. This produces almost identical primary components, and in the case of the lowest ionisation simulation a second component, similar to the one found in the neutral case (Fig.~\ref{varyingabs1}), is also returned. We find that as the ionisation parameter is increased, the strength of the component returned (in terms of the fraction of the spectral variability attributable to this component, and not to noise) lowers. This is due to the decreased effect of the absorption, and means that we are most likely to be able to detect neutral absorption variability using this method.

\section{PCA Results from the Deep XMM-Newton Sample}
\label{resultssection}
In this section, we present the results from PCA of the 26 sources in our sample. For each source, we show all the significant PCs returned and the fraction of the total variability attributable to each component. As in the case of the simulations presented earlier, the resulting component spectra are equally valid with the y axis inverted. Based on the PCA results obtained, we attempt to arrange and categorize the sources analysed in a logical manner, through comparison with the simulations presented in the previous section, and break the sample into four subgroups:
\begin{itemize}
\item The first source analysed using this method was MCG--6-30-15, which was presented in \citet{Parker14a}. This was found to have four significant principal components, although the fourth was weak and was not investigated in detail.
Having analysed our sample of AGN, we find four more sources (NGC~4051, 1H0707-495, Mrk~766 and NGC~3516) which display the same pattern of variability with four components, and several others which are limited by flux or lack of variability, but which show at least the first two components, with a similar spectral shape. The key features of this group of objects are the suppression of the continuum variability around the energies of the iron line an soft excess and a PC showing a strong correlation between these energy bands.
\item We find a second group of four objects (Ark~564, PKS~0558-504, Mrk~335 and 1H~0419-577) with a similar but qualitatively different pattern of variability. These sources show similar suppression of the primary powerlaw by a reflection component, but the pivoting term steepens sharply with energy.
\item In \S~\ref{absorbedsection} we discuss the three sources (NGC~1365, NGC~4151, and NGC~4395) that show good evidence of variable partial covering absorption. The higher order terms differ between these objects, and may indicate the presence or lack of intrinsic variability.
\item Finally, we include the PC spectra for the nine sources which do not appear to fit into the groupings discussed so far. These objects are presumably exhibiting different variability mechanisms, and we discuss them on an individual basis.
\end{itemize}

\subsection{Group 1: MCG--6-30-15 Analogues}
\label{mcg6objects}
Within the sample of objects we have analysed, four additional NLS1 sources that show the same four variable components as MCG--06-30-15 have been identified. These sources are NGC~4051, 1H0707-495, NGC~3516, and Mrk~766. We also found several sources which could be displaying the same variability pattern, but have lower data quality, making it impossible to be certain.

\subsubsection{MCG--6-30-15}

MCG--6-30-15 is very well studied, bright and highly variable narrow line Seyfert 1 (NLS1) galaxy. It was the first AGN in which a relativistically broadened iron line was found \citep{Tanaka95}, and also shows the characteristic features of warm absorption \citep{Otani96}

\begin{figure}
\centering
\includegraphics[width=7cm]{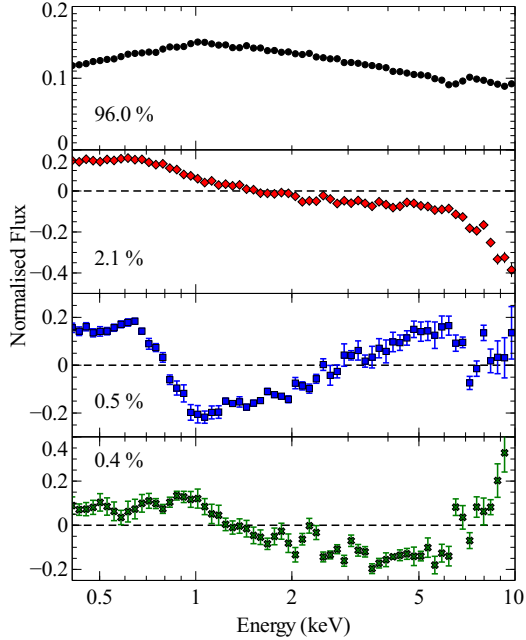}
\caption{Principal component spectra found in MCG--6-30-15, ordered from top to bottom by the fraction of variability in each component, as presented in P14a. The percentage of variability in each component is also shown.}
\label{mcg6_spectra}
\end{figure}

Fig.\ref{mcg6_spectra} shows the three PC spectra presented by P14a, along with the weak fourth component not discussed in that work. The first three PCs found in this object were analysed in detail by P14a, who found that they were well explained by the effects of a powerlaw varying in normalisation and photon index, and uncorrelated  variations in a relatively constant reflection spectrum. These findings are consistent with the light-bending interpretation of the variability in this AGN, in which the height of the primary X-ray source above the disk changes \citep{Martocchia00,Miniutti03,Miniutti04} leading to more extreme variations in the primary emission than in the reflected emission.

The suppression of the primary component at the energies of the soft excess and iron line indicate the presence of a strong, relatively constant spectral component at these energies. Likewise, the breaks in the second component correspond to the same energies, where the primary power law, and hence variations from the changes in photon index, are suppressed. This is best explained by a strong relativistically blurred reflection component, which is relatively constant when compared to the power law due to light bending. A partially covered power law can reproduce the spectrum of the source, but not the spectral shape of the observed PCs (P14a,b). In P14b we showed that the first three components could be produced by the variable continuum and reflection model, and that the predictions for either ionised or neutral partial covering absorption variability are completely different from those due to intrinsic variability (Fig.~\ref{varyingabs1},\ref{zxipcf_fig}) and therefore cannot explain the observed variability without extreme fine-tuning of multiple spectral components.

It is interesting that the reflection component shows a turnover below $\sim0.7$~keV (third panel), whereas the suppression of the continuum variability (first panel) shows no such break. This suggests that the variable reflection component isolated here is not the sole origin of the soft excess in this source. The fourth order component also appears to contribute to the soft excess, although it is strongly suppressed by the reflection component. The origin of this component remains a mystery, as we have so far been unable to convincingly reproduce it using simulations.

\subsubsection{NGC 4051}

NGC~4051 is a NLS1 which extremely variable on all time scales. \citet{Ponti06} showed that the spectrum of the source in various flux states could be well described by a power law plus relativistic reflection model, like that used in MCG--6-30-15. It also exhibits strongly flux-dependent time lags \citep{Alston13}, which favour models involving intrinsic variability and relativistic reflection over reprocessing by distant material.

\begin{figure}
\centering
\includegraphics[width=7cm]{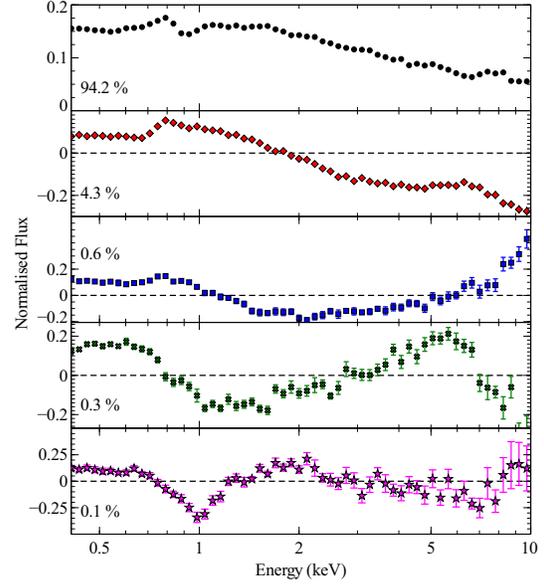}
\caption{Principal component spectra found in NGC~4051, ordered from top to bottom by the fraction of variability in each component. PCs one, two, three and four correspond to one, two and four and three from MCG--06-30-15, and the additional component 5 appears to be attributable to a variable absorption feature.}
\label{4051_spectra}
\end{figure}

Our analysis of NGC~4051 reveals five significant PCs, shown in Fig.~\ref{4051_spectra}, four of which correspond well to those found in MCG--6-30-15. PCs one, two and four match components one, two and three from MCG--06-30-15 and the simulation shown in Fig.~\ref{reflectionsims}. These components show the same breaks and dips, and differ only quantitatively. The third PC from NGC~4051 appears to match the weak fourth component found in MCG--6-30-15. Finally, the fifth component, which has no analogue in MCG--6-30-15, shows what appears to be an absorption edge at an energy of $\sim$1~keV, with no other strong features visible in the spectrum. We suggest that this component corresponds to a change in the properties of an ionized absorber, as described in \citep{Ogle04}, however this conclusion is extremely tentative and must be treated with caution because of the high order of this PC. We conclude, based on the almost identical components produced, that the variability in this source is driven by the same processes as in MCG--6-30-15.

We note that the variability is dominated by the first (power law) component, the spectrum of which is qualitatively very similar to the RMS spectrum of NGC~4051 presented by \citeauthor{Ponti06}. This is unsurprising, given that both methods should produce a spectrum of the relative strength of variability (which is largely due to the variable powerlaw in both cases) but it is an interesting confirmation of our method.

The much higher signal to noise in the third component relative to the fourth component in MCG--06-30-15 gives us the opportunity to investigate this component in more detail, and hopefully to understand its origin. In \S~\ref{simsection}, we showed that a component with this spectral shape can be produced by adding a second reflection component to the spectrum, with a higher ionisation parameter and more extreme relativistic blurring. 

\subsubsection{1H0707-495, NGC 3516 and Mrk 766}

\begin{figure*}
\centering
\includegraphics[width=14cm]{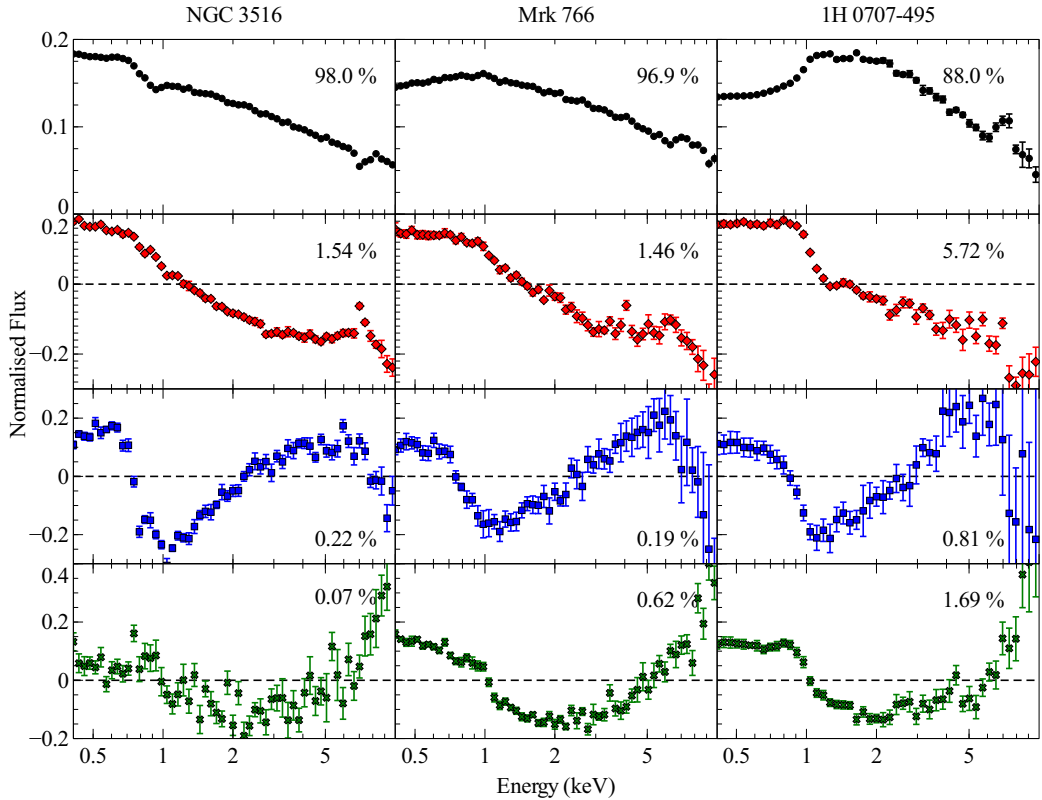}
\caption{Principal component spectra found in NGC~3516 (left), Mrk~766 (middle) and 1H~0707-495 (right). The spectra are ordered from top to bottom, with the exception of PCs 3 and 4 in Mrk~766 and 1H~0707-495, which have been swapped for ease of comparison. These components are qualitatively the same as those found in MCG--06-30-15 and NGC~4051.}
\label{3516and766}
\end{figure*}

1H~0707-495, NGC~3516 and Mrk~766 all display the same four variable components as MCG--06-30-15 and NGC~4051, again with the only large difference being the order of the third and fourth component, which is reversed with respect to MCG--06-30-15 in 1H~0707-495 and Mrk~766. 
The component spectra of these objects are shown in Fig.~\ref{3516and766}, with the third and fourth components swapped in Mrk~766 and 1H0707-495 for ease of comparison. All four components are significantly detected, and appear to be almost identical to those found in the previous sources.

The similarity of the PCs obtained from these three sources suggests that the same physical processes are causing the majority of the variability in each source, with only quantitative differences between them, such as the relative strengths of the components or the exact parameters of the relativistic reflection spectrum. We note that, in all the sources discussed so far, the drops in the primary component at low energies and around 6~keV correspond extremely well to the shape of the component attributed to reflection variability by P14a. This is very good evidence for the presence of relativistic reflection in these objects. While warm absorption features are clearly visible in the spectra of some of these objects, we find no evidence of spectral variability clearly attributable to absorption in the PCs returned. We discuss this further in \S~\ref{abssims} and \S~\ref{discusswarmabs}.

The broad iron line features in the third or fourth components of all 5 sources so far discussed appear to peak slightly below 6.4~keV, and are generally slightly misaligned with the suppression features in the first and second components. There are several factors which are likely to contribute to this. Firstly, we must account for presence of distant reflection in these objects. With the exception of 1H~0707-495, all these sources have narrow iron lines, distinct from the broad component. Because distant reflection can be regarded as constant on these time scales the effect of this emission is to strongly suppress the variability at 6.4~keV (see \S\ref{section_distref}). This is particularly obvious in the first two components of NGC~3516, which has a very strong narrow line \citep[e.g.][]{Markowitz08}. This suppression will also affect the blurred reflection component, suppressing the 6.4~keV variability and therefore shifting the broad line peak to lower energies. Secondly, as the red wing of the line originates closer to the event black hole, it should be more rapidly variable than the blue edge of the line \citep[See e.g. the frequency resolved iron~K lags in ][]{Zoghbi12,Kara14}. This will skew the PC line profile towards the most variable part, rather than reflecting the true shape of the line. Finally, we stress that we have deliberately chosen conservative parameters for the relativistic blurring in \S\ref{refsimssection}, with an emissivity index of 3 whereas many, if not all, observed sources have steeper profiles \citep{Walton13_2}.

The presence of a separate reflection component in the PCA results is clear evidence of at least a partial disconnect between the reflected and the continuum emission. We note that the 10~ks intervals used for this analysis is considerably longer than the light travel time over the inner disk (on the order of tens to hundreds of seconds), and hence reverberation lags, for these objects. We therefore suggest that the variations found here could due to changes in the accretion disk, particularly in the ionisation parameter, which can take place on longer time scales and introduce independent reflection variations. Alternatively, this behaviour could be symptomatic of light bending effects, which can suppress the variability of the reflection component \citep[e.g.][]{Miniutti04}.

\subsubsection{Other objects}

We show here the five remaining objects that show the same (or very similar) first two components, with suppression of the first component at low and high energies and flattening of the second above 2 keV. The conclusion that these objects are showing the same variability pattern is weaker, due to the lower quality or less variable data used, so some of these objects may yet have a different physical origin for their variability. Nevertheless, our current analysis is consistent with the same behaviour

The objects in this subgroup are IRAS~13224-3809, PG~1211+143, NGC~2992, MCG-5-23-16, and NGC~5506. The two components from all these sources are shown in Fig.~\ref{nearlymcg6_1} and Fig.~\ref{nearlymcg6_2}.

\subsubsection{IRAS~13224-3809 and PG~1211+143}
\begin{figure}
\centering
\includegraphics[width=8.5cm]{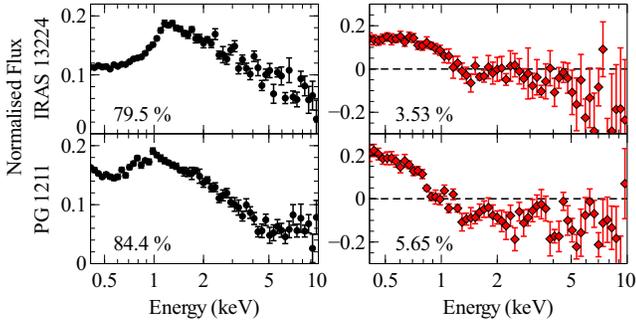}
\caption{First (left) and second (right) PCs from IRAS~13224-3809 and PG~1211+143. The primary component found in both of these sources is very simlar to that found in 1H~0707-495 (Fig.~\ref{3516and766}), and the second components show some degree of flattening at high energies.}
\label{nearlymcg6_1}
\end{figure}

IRAS~13224-3809 and PG~1211+143 both show a sharp break at around 1~keV in their first PC spectra (Fig.~\ref{nearlymcg6_1}), which is almost identical to that found in 1H~0707-495 (Fig.~\ref{3516and766}).  The second components of both objects appear to be also very similar, but both are heavily degraded by noise at high energies. We therefore tentatively group them with the others in this class
We note that PG~1211+143 is around an order of magnitude more massive than 1H~0707-495 \citep{Kaspi00,Peterson04}, and its reverberation lags appear to be correspondingly shifted to longer frequencies and larger amplitudes, but otherwise very similar \citep{DeMarco11}. 

\subsubsection{NGC~2992, MCG-5-23-16, and NGC~5506}
\begin{figure}
\centering
\includegraphics[width=8cm]{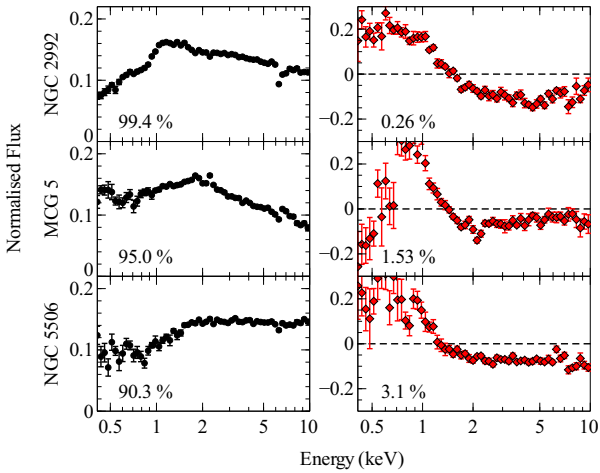}
\caption{First (left) and second (right) principal components from NGC~2992, MCG-5-23-16, and NGC~5506. Unlike the other sources in Class A, these objects show a lot more noise at low energies, due to the effects of neutral absorption. However, the components returned appear to be broadly similar to those found in the unabsorbed objects.}
\label{nearlymcg6_2}
\end{figure}

These three sources are qualitatively different from the other objects in this class. They show no clear signs of a break at high energies in the second component, and the quality of both components returned is noticeably worse at low energies (Fig.~\ref{nearlymcg6_2}). In fact, these objects are Compton thin Seyfert~2 sources, so it is possible that the resemblance between them and the other objects in this class is only superficial.

The shape of the first component in all three sources, with a break at low energies and a suppression feature at $\sim$6.4~keV suggests that this component corresponds to the intrinsic source variability, rather than absorption variability, which would have no effect at high energies given the column density in these AGN. The low energy drop then corresponds to the soft excess (some of this may also be from diffuse thermal emission, as in NGC~1365), and the 6.4~keV feature corresponds to the iron line produced by distant reflection.

The second component is more ambiguous, and could potentially be produced by pivoting of the primary power law (as in the sources described above), changes in the column density of the absorption (as in NGC~4395, NGC~4151 and NGC~1365, see \S~\ref{otherthings}), slow variations of a soft excess component, or a combination of mechanisms.

\subsection{Group 2: Ark~564 Analogues}
\label{Ark564section}

\begin{figure*}
\centering
\includegraphics[width=16cm]{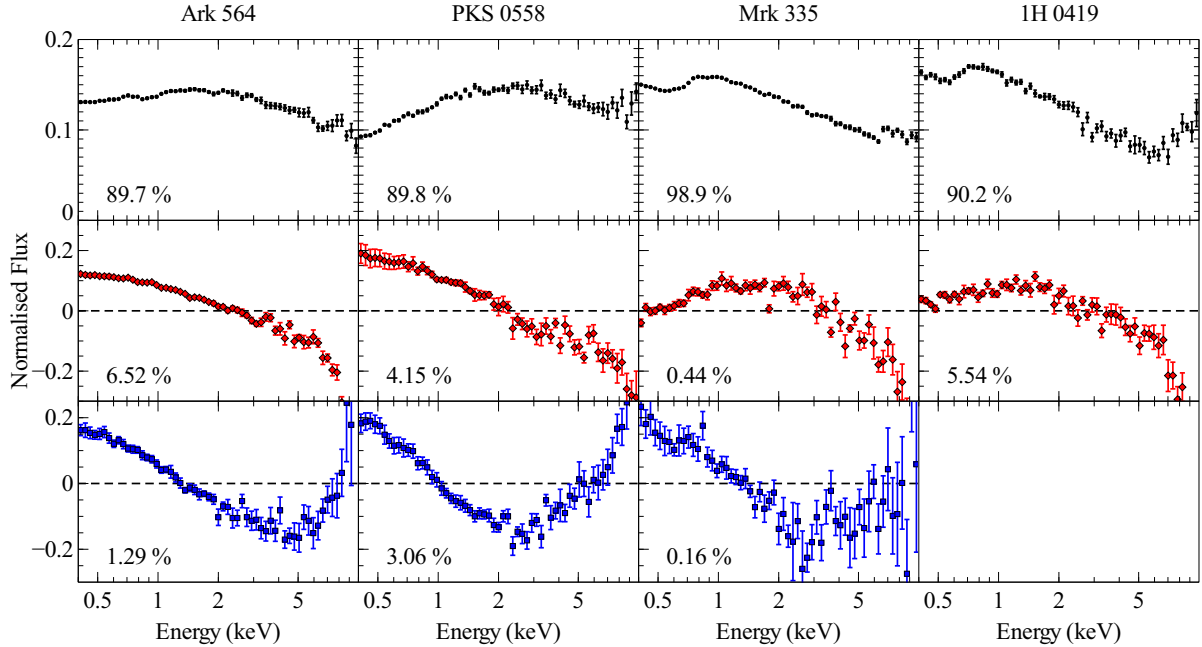}
\caption{First (left, black), second (middle, red) and third (right, blue) principal components from the sources which show a steepening with energy of the second component, and some suppression of the primary component at low energies. Three of these sources also show a third significant component, which has roughly the same shape in all objects.}
\label{nearlyark564}
\end{figure*}

This second major class includes all the objects where the second (pivoting) PC steepens with energy, rather than flattening like those described in \S~\ref{mcg6objects}, and which also show some low energy suppression of the first component. These sources are Ark~564, PKS~0558-504, Mrk~335, MR~2251-178 and 1H~0419-577. 
There is much more variety in the shape of the components (particularly the primary components) found in these objects than there is in the Class~A objects, which may indicate that there are several different variability mechanisms in these objects. The PCA spectra for these objects are shown in Fig.~\ref{nearlyark564}.

\subsubsection{Ark 564}

The first PC found for Ark~564 shows features that are very similar to those found in the primary components of the Class A objects: suppression of the power law variability at low energies, and at the energy of the iron line. However, these features are rather weaker than in the Class~A sources, implying that the constant component giving rise to them is correspondingly weaker. The second component is also much smoother, flattening at low energies and steepening at high energies, with no sharp breaks. Finally, the third component shows a correlation between low and high energies, but with no noticeable iron line feature. This is qualitatively similar to the third or fourth order component found in the objects shown in Figs.~\ref{mcg6_spectra}, \ref{4051_spectra} and \ref{3516and766}, and can be produced by the presence of a variable high ionisation reflection spectrum, as shown in Fig.~\ref{reflectionsims}.

A Comptonization or Bremsstrahlung model with no reflection for the constant component of the spectrum does not extend to high enough energies to explain the observed curvature in the second and third components, and cannot produce the third PC at all, which requires an excess at both low and high energies. As in the earlier sources, we suggest that this component can be identified with the soft excess, and the high energy upturn suggests a reflection origin. The lack of an iron line feature in this component then implies either suppression from another, less variable reflection component (such as distant reflection) or that the reflection is highly ionised. Alternatively, there could be a correlation between the soft excess component and another, harder component, causing the high energy variability. We discuss this further in \S~\ref{discusscomp4}.

\subsubsection{PKS 0558-504}

PKS 0558-504 shows very similar variability to Ark~564, producing three components which show the same features. The first PC is suppressed at both low and high energies, and the second shows the same steepening with flux. The minimum of the third component appears to be at a lower energy than in Ark~564, but it shows the same correlation between low and high energy bins.

This object is radio loud, which indicates the presence of a jet. However, its spectrum and variability appear to be fairly standard for a NLS1 \citep{papadakis10}.

\subsubsection{Mrk 335 and 1H 0419-577}

These two sources show clear suppression of the primary PC at low energies and at around 6~keV, and the second component converges towards zero at low energies in both objects. 
As in the case of the Class~A objects, the shape of the primary components returned for these objects cannot be explained by a variable absorption model, requiring a constant component at low energies and at 6 keV which is strong enough to cause the observed spectral features (see \S~\ref{simsection}).

There is a hint of structure in the third component of Mrk~335, although it is noisy, and such structure is noticeably absent from the other sources in this group. Combined with the heavy suppression of the pivoting component at low energies, this suggests that these two objects may be qualitatively different from Ark~564 and PKS~0558. One possible issue with the PCA of these objects is that they have been observed in with \emph{XMM-Newton} in both high and low flux states, which may mean that the components returned are dominated by variability between, rather than during, the observations. 

\subsection{Group 3: Variable Absorption Sources}
\label{absorbedsection}

We present here the three objects which show clear evidence of strongly varying partial covering neutral absorption. These objects are NGC~4395, NGC~1365 and NGC~4151. NGC~1365 was analysed in detail in P14b, and we discuss this more below.

\begin{figure*}
\centering
\includegraphics[width=11cm]{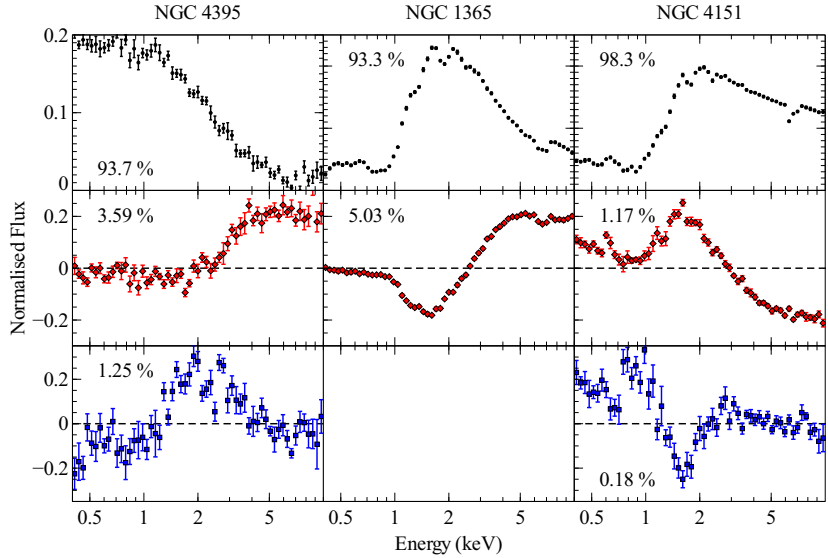}
\caption{First (left, black), second (middle, red) and third (right, blue) principal components from the sources which do not appear to fit into either of the categories defined in \S~\ref{mcg6objects} and \ref{Ark564section}. Third components are only shown if they are found to be significant. Objects are grouped with those similar in appearance.}
\label{absorbedthings}
\end{figure*}

\subsubsection{NGC 4395}
\label{sec_ngc4395}
The variability in NGC~4395 is dominated by absorption \citep{Nardini11}. The first two components produced here are an almost perfect match to the simulations of a partially covered power law, where the covering fraction and intrinsic source flux are both allowed to vary (see \S~\ref{simsection}). We suggest that the weak third component is caused by changes in the column density, which requires a correction factor be added to the absorption term (see component 2 in Fig.~\ref{varyingabs1}.

We note that this source does show evidence of reflection features in the spectrum and a reverberation lag \citep{DeMarco13}, but any reflection features in the PCA components are completely swamped by the
absorption variability.

\subsubsection{NGC 1365 and NGC 4151}
These two sources are very similar in terms of their behaviour. They both show thermal emission from gas around the nucleus, which has been resolved with \emph{Chandra}. This emission dominates the spectrum below $\sim 2$ keV, and almost completely damps out the AGN variability. This is why the first and second components of both sources are strongly suppressed at low energies.

We recently analysed the \emph{XMM-Newton} data on NGC~1365 using PCA, (P14b) and we refer the reader to that work for detailed explanation of the results. The main conclusions were that the variability in NGC~1365 is dominated by changes in the column density and covering fraction of the absorber, but intrinsic variability can also be distinguished when considering only relatively unobscured observations.

\subsection{Group 4: Other Sources}
\label{section_otherthings}
\begin{figure*}
\centering
\includegraphics[width=15cm]{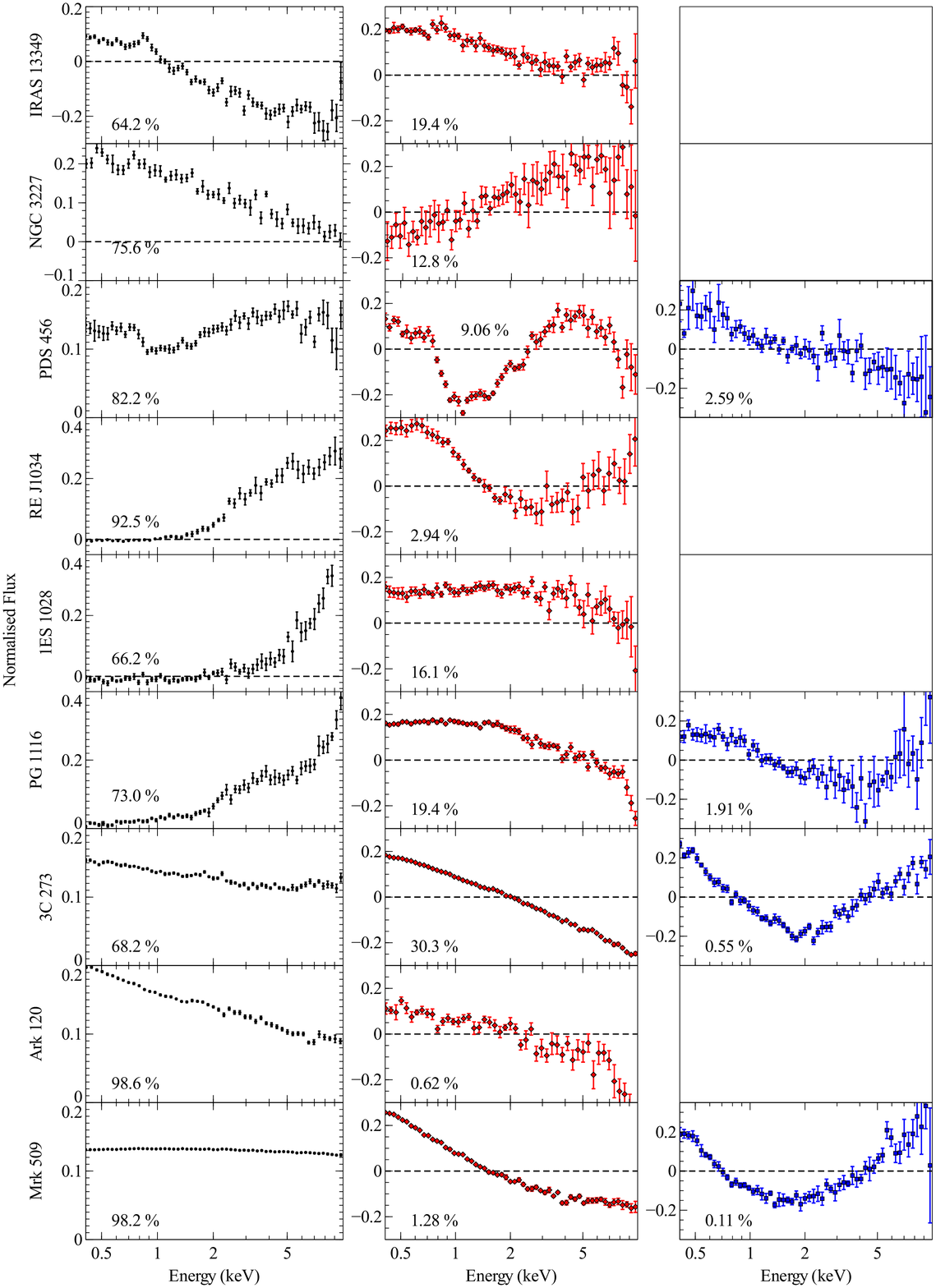}
\caption{First (left, black), second (middle, red) and third (right, blue) principal components from the sources which do not appear to fit into either of the categories defined in \S~\ref{mcg6objects} and \ref{Ark564section}. Third components are only shown if they are found to be significant. Objects are grouped with those similar in appearance (firstly by primary component, then secondary). The fractional variability of each PC is also shown.}
\label{otherthings}
\end{figure*}

All the objects which do not appear to fit into the classes defined above are shown in Fig.~\ref{otherthings}, grouped with those that appear to be similar, if any. In this section we will briefly discuss the variability patterns in each of these objects, and potential origins for the PCs shown here.

\subsubsection{IRAS 13349+2438}
This source the only source in our sample where the primary component shows a significant anticorrelation between low and high energies. This first term appears to be qualitatively very similar to the pivoting terms found in the objects described in \S~\ref{mcg6objects}, showing the break at $\sim 1$ keV, with a possible second break around 7 keV. However, there is no PC corresponding to the normalisation of the primary emission. The second term is consistent with being entirely positive, and is largely suppressed at high energies. This is very similar to the first order term found in NGC~4395, and the component produced in simulations of an absorber that varies in covering fraction (see \S~\ref{simsection}) and indeed this source does show a strong warm absorber \citep{Sako01}. IRAS~13349+2438 also shows a marginally significant soft lag \citep[90\% confidence,][]{DeMarco13}, which supports the interpretation of the first component as a pivoting power law component in the presence of relativistic reflection, however the absence of detected variability in the primary emission remains a mystery.

\subsubsection{NGC~3227}
A recent analysis of the variability in the Seyfert 1.5 NGC~3227 by \citet{Arevalo14} showed the spectral variations could be described by a two component soft excess plus power law model, where both components vary independently. This is consistent with the separate soft and hard components returned by our analysis, although the signal is weak (particularly in the second component). However, this object does appear to be particularly unusual (or the state in which it was observed was unusual), so these conclusions may not be universally applicable.

An alternative model was proposed by \citep{Noda14}, where the variability arises from multiple power law components, one hard and one soft, with a constant distant reflection component. It is possible that such an arrangement of power laws could reproduce the observed PCs, and there are some similarities between the 2nd PC found here and that simulated for the two power law model in Fig.~\ref{sims_twopls}.

\subsubsection{PDS 456}
The primary component we find for the $z=0.184$ quasar PDS~456 appears to be unique. Rather than being suppressed at the energy of the soft excess and iron line, this component appears to be enhanced. The second component appears to be a good match for the third or fourth order components described in \S~\ref{mcg6objects} (although more redshifted), which are attributed to relativistic reflection variability. PDS~456 is rapidly variable and shows a strong absorption edge at $\sim8$~keV \citep{Reeves00}, and there is strong evidence of high-velocity outflowing material \citep{OBrien05,Reeves14}. The broad band spectrum has been successfully modelled using a standard relativistic model by \citet{Walton10}.

It is surprisingly difficult to reproduce a PC with the spectral shape of the primary component of PDS~456. Allowing the reflected emission to scale with the primary power law does not work, as the fractional deviation produced by this is the same as in the case with no reflection at all. The same problem applies to a constant absorption component. Allowing for a variable partial covering absorber produces no variability at high energies, and while a variable reflection component alone could produce a similar component to this, it does not make intuitive sense to have the reflected emission more variable than the primary power law emission.
One possible way that this component could arise is in the case where the primary emission, modified by ionised absorption, varies with respect to an unabsorbed component (i.e. variable partial covering ionized absorption, but more complex than the simple case considered in \S~\ref{abssims}). In this case, the spectral features imposed on the power law by the absorption are preserved in the PC spectra.

\subsubsection{RE J1034+396}
The PC spectra for this source show that the primary variable component, presumably the power law emission, is heavily suppressed at low energies,with a second component that shows a strong excess at low energies and a possible increase again at high energies.

\subsubsection{1ES 1028+511 and PG 1116+215}
The primary component in both of these sources is completely suppressed at low energies, similar to that in RE~1034+396. However, the second PC is much more like those found in Ark~564 and PKS~0558-504, showing the same steepening at high energies.

\subsubsection{3C 273}
The first two components returned from the analysis of 3C~273 are the most featureless we find for any source, consistent with an almost completely power law dominated spectrum. There is a slight decrease with energy in the primary component, which suggests the presence of a small relatively constant component in the spectrum. We note that in this object the second order pivoting term makes up a much larger fraction of the variability ($\sim 30$ per cent) than in the other sources that show a significant pivoting term ($\sim 5$ per cent). This is probably due to the emission from the jet, which changes significantly in photon index.

The third order term appears to introduce a break in the spectrum at around 2~keV, which is well constrained and much sharper than the third order PCs seen in other sources. This could be due to the interplay of the two power law components from the jet and corona \citep[e.g.][]{Pietrini08}.

\subsubsection{Ark 120}
We find an extremely straight primary component from the PCA of Ark 120, which is unexpected given the extremely large soft excess shown by this source. This implies that the origin of the soft excess must be strongly correlated with the primary power law emission, and suggests a different origin for the soft excess in this source and the sources in \S~\ref{mcg6objects} and \ref{Ark564section}. The gradual decrease with energy implies, as in 3C~273, that there is a relatively constant component which increases with flux but without any strong broad features. This could potentially be explained by a distant, neutral reflection component, and we note the possible narrow features around 6.4~keV in the primary PC. 

\citet{Nardini11_2} found that the \emph{Suzaku} spectrum of Ark~120 was well described by a blurred reflection model. However, \citet{Matt14} found that a joint \emph{XMM-Newton} and \emph{NuSTAR} observation could be modelled with the \textsc{optxagn} Comptonisation model \citep{Done12} as well as distant reflection, while models with a relativistic reflection component were ruled out. It may be that the soft excess in this source is dominated by Comptonisation strongly correlated with the continuum, while the earlier sources have a larger contribution to the soft excess from reflection.

The second component shows a steepening at high energies, much like the second components of the objects in \S~\ref{Ark564section}. This probably indicates the presence of a distant reflection component, where the iron line and absorption edge around 7~keV cause the break in the pivoting term (see \S~\ref{simsection} for more discussion on this).

\subsubsection{Mrk 509}
The three components returned for Mrk~509 are qualitatively different from those found for the objects discussed above. There is almost no curvature in the first component, and no obvious breaks in the second, implying that the  strong and relatively constant component found in the other objects is not present here. The third component is also significantly different. It shows a much weaker iron line feature, and the spectral breaks are much less pronounced. We suggest that this component can be identified with the neutral reflection discussed in \citet{Ponti13}, as the strength of the iron line feature is strongly dependent on the ionisation parameter.

Both the second and third order terms in Mrk~509 are very weak, when compared with the other objects in this class. We note that this source is relatively massive \citep{Peterson04}, and is therefore less variable.

\subsection{Summary}
Applying PCA to our sample of AGN has revealed a large number of principal components, many of which match well to the simulated components in \S~\ref{simsection}, although many more remain unidentified. We have successfully identified four sources which show clear evidence of relativistic reflection, and three which show clear evidence of cold absorption. A further four sources are consistent with having the same, so far poorly understood, pattern of variability, and a final nine sources show variability patterns that differ strongly from both the major groups of sources and from each other. 

\section{Discussion}
\label{discussion}

\subsection{Additional discussion of selected sources}
\emph{MCG--6-30-15:} 
We note that MCG--6-30-15 and all of the other analogous sources where four components are found exhibit lags between the primary emission and the soft excess at $>99\%$ confidence \citep{DeMarco13,Emman11}, which correlates strongly with the black hole mass. This is interpreted as evidence of a delay between the primary emission and the reflected emission from the disk, and in all cases the lag is on the same order as the light travel time for a few gravitational radii.
An alternative model was proposed for spectral shape and variability of MCG--6-30-15 by \citep{Miller08}, who suggested that the red wing of the iron line could instead be produced by a combination of ionised absorbers. However, this model has some serious flaws. It requires that the absorption be strongly correlated with the strength of the primary emission \citep{Reynolds09} to explain the relative constancy of the iron line feature; it cannot explain the strong correlation between the soft excess and the iron line, as these arise from different components (P14a); an absorption model cannot produce the negative time lags seen in this and other similar sources \citep{Emman11}; and it is disfavoured by broad-band spectral fitting \citep{Marinucci14}. 

\emph{1H 0707-495:} This is a well known source which displays strong evidence of prominent relativistic reflection. A 500~ks observation with \emph{XMM-Newton} in 2008 revealed the presence of both iron K and L emission lines, and a reverberation lag of around 30~s between the continuum and reflection emission. A more detailed examination of this lag by \citet{Kara13_1h0707}, using over 1~Ms of data, found evidence of an iron K line in the lag-energy spectrum. In general, the spectral shape and variability of 1H~0707-495 can be modelled with a power law continuum plus two separate relativistic reflection components, with the same blurring parameters but different ionisation states \citep{Zoghbi10, Dauser12, Fabian12}. This is suggested to correspond to regions of different density on the surface of the disk, potentially caused by turbulence. Such a highly ionised reflection component, in the presence of a more constant and less ionised component, could potentially produce the PCs shown in the bottom panel of Fig.~\ref{3516and766}, but detailed modelling of this is beyond the scope of this work.

\emph{NGC~3516: } NGC~3516 shows several zones of warm absorption, described by \citet{Medhipour10}, who showed that variations in the covering fraction of the warm absorbers cannot be solely responsible for large variations in the flux, and therefore that intrinsic source variability plays a large part in the the spectral variability of the source. \citet{Turner11} argue that the lack of reverberation lags in this source is evidence that the flux variations are not intrinsic to the source, however \citet{DeMarco13} found such a lag at $>99\%$ confidence, indicating reflection from the accretion disk and the presence of intrinsic variability. 
\citet{Turner02} found evidence of a relativistically broadened iron line in NGC~3516, and \citet{Markowitz08} found that this was still required after complex absorption is taken into account, and adding an additional absorbing component could not reproduce the spectral curvature.
We note that some of the strongest variability, potentially due to an extreme absorption event \citep{Turner11}, is seen in the 2005 \emph{Suzaku} spectrum of this source, and will therefore not manifest in our results.

\emph{Mrk 766:}
The origin of the spectral shape in Mrk~766 is controversial. \citet{Page01} found that in a 60~ks \emph{XMM-Newton} observation the spectral variability could be explained by a powerlaw plus relativistic reflection model, however \citet{Miller07} and \citet{Turner07} analyse a longer, 500~ks observation and argue that the variability is instead better explained by variable absorption. More recently, \citet{Emman11} found almost identical time lags in Mrk~766 and MCG--6-30-15, indicating reverberation close to the black hole.

\emph{IRAS 13224-3809:} 
This object is very similar to 1H~0707-495 in both spectral and timing properties, with a strong soft excess and turnover around 7~keV \citep{Boller02,Boller03}. A 500~ks observation of this source with \emph{XMM-Newton} was analysed by \citet{Fabian13}, who found evidence for relativistically smeared iron K and L lines, high spin, and a reverberation lag of 100~s. \citet{Kara13_iras} found that the reverberation lags were dependent on the source flux, with the lag shifting to higher frequencies and smaller amplitudes as the source flux drops, consistent with the light bending model, where the corona is closer to the black hole in low flux states.

\emph{Ark 564:}
Ark~564 has been found to show both a highly significant soft lag \citep{DeMarco13}, and an iron K lag \citep{Kara13}, which strongly indicates the presence of relativistically blurred reflected emission in this objects.

\emph{Mrk~335 and 1H 0419-577:}
Like Ark~564, Mrk~335 shows both a soft lag and an iron K lag \citep{DeMarco13,Kara13}. We note that both of these sources have been observed in extremely low flux states \citep{Grupe08,Pounds04}, which have in the past been interpreted in terms of both relativistic reflection \citep{Fabian05,Gallo13} and absorption models \citep{Pounds04b,Turner09}, which give equivalent fits. Analysis of variability in these sources can break the deadlock between the two models, for example, the discovery of a time lag between the iron line and the continuum emission in Mrk~335 by \citeauthor{Kara13} is very strong evidence of relativistic reflection in this object, and a recent analysis of spectral variability with \emph{NuSTAR} \citep{Parker14c} in a low state of the same source found that the variability could be well explained using simple light-bending models. The \emph{NuSTAR} spectrum also shows an extremely strong broad iron line and Compton hump, while the low energy spectrum was observed simultaneously with \emph{Suzaku} and is also well modelled by reflection \citep{Gallo14}

\emph{NGC 1365 and NGC 4151:}
The recent joint \emph{NuSTAR} and \emph{XMM-Newton} analyses of the spectrum \citep{Risaliti13} and the variability (Walton et al., submitted) of NGC~1365 show clear evidence of relativistic reflection in this source, stretching up to 80~keV.
NGC~4151 shows strong evidence of reverberation in the iron K band \citep{Zoghbi12,Cackett13}, again indicating the presence of relativistic reflection, and has also recently been observed with \emph{NuSTAR}, again showing a broad line and Compton hump (Keck et al., in preparation).

\emph{PDS 456:}
The spectral variability of this object was examined extensively by \citet{Behar10}, who found that it exhibits both variable reflection and absorption, and modelled the spectral variability using an unabsorbed reflection component and a partially covered absorbed powerlaw, where the reflection spectrum originates from the outflow and is thus unabsorbed and blueshifted.
We also note that PDS~456 is likely to be an extremely massive object, on the order of $10^9 M_\odot$ \citep{Zhou05}, and this means that the reverberation timescale could easily be greater than the 10~ks intervals used for this analysis. This could potentially affect the observed relationship between the continuum and reflected emission, causing different components to be found for this source than for other, less massive, objects.

\emph{RE J1034+396:}
This is an extremely unusual AGN, in that it has shown a highly significant quasi-periodic oscillation (QPO) \citep{Gierlinski08}. The QPO arises from the power law continuum, which dominates the variability at high energies, as shown by \citet{Middleton09} who concluded that the soft excess was most likely to be caused by low-temperature Comptonisation. The QPO was identified in 5 additional observations by \citet{Alston14}, who found that it was only significantly detected in the hard band. It therefore seems likely that the QPO is associated with the hard component we identify here, which dominates the spectral variability. In addition, the recent detection of a significant QPO in the AGN MS~2254.9-3712 by \citet{Alston14_ms2254}, which shows very similar PCs to RE~J1034+396, raising the intriguing possibility that this combination of PCs indicates the likely presence of a QPO.
The variability in this source was also investigated by \citet{Zoghbi12}, who found that the spectrum could be modelled using a low temperature black body, a strong and highly ionised blurred reflection component which dominates below around 2~keV and a power law. \citet{Zoghbi12} also found an energy dependent lag of around 100~s in one of the observations (obs. ID 0506440101), indicating a delay between the primary and reflected emission.
We note that it is entirely possible that both Comptonisation and relativistic reflection are present in the source spectrum and contribute to the soft excess, thus explaining the observed variability and lag.

\subsection{Origin of the 4th MCG--6-30-15 component}
\label{discusscomp4}
The fourth PC in MCG--6-30-15 and NGC~3516 and third PC in NGC~4051, Mrk~766 and 1H~0707-495 is still poorly understood. We identify three possibilities for its origin: firstly, and most simply, this PC could represent an independent physical component; secondly, it could arise as a correction factor between two spectrally similar physical components (as in the black body simulations in \S~\ref{bbsims}); finally, it could represent a more complex interaction between the spectral components or a change in their parameters.

We have so far been unable to reproduce this component through simple models of spectral variability, but some conclusions can be drawn from the shape and variability. One notable feature is that the shape of the component does not appear to change greatly with the ordering of the components - it is not systematically different between those objects where it is the third PC and those where it is fourth. This suggests that this component is not a correction factor to the lower order components (as this should cause changes in the shape with the ordering of the components), and instead represents real physical variability. 
On a related note, the ordering seems to change with the time scales probed. When we use 5~ks instead of 10~ks spectra for the PCA of MCG--6-30-15, the ordering of the third and fourth components switches, as do those in NGC~4051 when we use 20~ks spectra. This suggests that the unidentified component is variable on shorter time scales than the reflection component.

The spectral shape of this component implies that it contributes at both low and high energies. It appears to be contributing to the soft excess, but there is a strong upturn above 5~keV. One possible explanation for this could be that a soft excess component is associated with changes in the continuum, but this remains speculation until we can successfully simulate this variability. This is qualitatively similar to the third components found in Ark~564 and PKS~0558 (and possibly Mrk~335), and may indicate a common origin.

\subsection{Redundancy in spectral components}
As discussed in \S~\ref{bbsims}, spectral components that have a similar effect on the total spectrum (such as pivoting of the power law and changes in the flux of a black body, which both cause changes in the spectral hardness) may not result in distinct PCs, rather there will be one dominant PC expressing the average variations and one `correction factor' PC which accounts for the differences between the two. It is therefore extremely important that we can distinguish between PCs that accurately represent spectral component and those which are averages or corrections. 

This is only really a problem for those components that we have not yet managed to simulate in detail, and only for components of second order or higher. Aside from the absorbed sources (where corrections have to be made for changes in the column density as the assumption of simple additive components breaks down) we are not aware of any components that are likely to be corrections due to similar spectral components, however this should be considered as a possible explanation for the poorly understood PCs, particularly in \S~\ref{section_otherthings}.

\subsection{Lack of Ionised Absorption Variability}
\label{discusswarmabs}

With the possible exception of the fifth PC in NGC~4051, we have not detected any clear evidence for strong variations in ionised absorption. This is not to say that there is no such variability - many of the sources studied here show clear evidence of ionised absorbers, and variations have been found in their properties between observations by many authors. Instead, we believe that the PCA method presented here is not optimal for the study of such variability. The simulations of variability in ionised absorbers presented in \S~\ref{abssims} returned components that were increasingly weak as the ionisation increased, and in real data this variability could easily become lost in the noise. Alternatively, we could simply be probing the spectra on the wrong time scales to pick up variability in the warm absorption, which should generally be operating at lower frequencies than the intrinsic changes.

It is possible that PCA could be used effectively to analyse variable absorption features using grating spectra, and it could potentially be used to identify correlated features and hence different absorption zones.

\subsection{Limitations and scope for future work}
A problem with this kind of analysis, where all available data from multiple observations often several years apart is combined lies in the implicit assumption that the nature of the source variability  has not changed qualitatively over the intervening period. In general this assumption appears to be valid - we have tested running the analysis on subsets of the data from several sources (MCG--6-30-15, NGC4051, Mrk~509, NGC~1365) where there is plentiful data, and found that this does not cause significant differences to the results. This holds whether we divide the observations by time, randomly, or by flux, although selecting by flux removes a lot of the variability and thus higher order terms. However, some sources have only two or three observations available (e.g. Ark 120), in which case we cannot rule out the possibility that we are looking at different variability mechanisms between observations. When looking in greater detail at individual sources or small samples we would ideally attempt to separate states where different variability processes are taking place. In this case, due to the large sample size and our desire to not bias the study by manually dividing the data into different states, we believe the optimum approach is simply to include all the data. The reader should therefore be aware that in some cases there may be multiple variability processes taking place within the same dataset, thus confusing the output components. However, in the vast majority of cases over 90 per cent of the variability is attributable to the first component, which implies that any changes in the nature of the variability are confined to higher order terms. We also note that even the longest time scales in this study ($\sim10$~years) correspond to only minutes to hours when scaled down to the masses of X-ray binaries. As state transitions in binaries take significantly longer than this, it seems reasonable to assume that the variability processes in AGN should be stable on these time scales.

There is great scope for extending this analysis, both by examining these sources in more detail and by looking at other objects. In this work we have restricted the analysis to the \emph{EPIC-pn} data, as it is of the highest quality. However, as briefly discussed in \citep{Parker14a}, it should be possible to combine data from different instruments using this method provided that it is handled carefully. By including the data from the MOS detectors in the analysis more details about the variability in these sources could potentially be discerned. There is also a wealth of data from other telescopes that has so far not been examined using this method. This could both expand the pool of sources and be combined with the \emph{XMM} data for more detailed examination of previously studied sources.

These is potential for using this technique with \emph{NuSTAR} \citep{Harrison13}. Although the relatively low count rate in AGN spectra mean that PCA will probably only be effective with a few of the brightest sources with long exposures, NuSTAR has revealed spectra of X-ray binaries of exceptional quality \citep{Tomsick13,Miller13b,Miller13a} which may be ideal for applying PCA to. The combination of hard and soft detectors on board \emph{ASTRO-H} \citep{Takahashi08} may allow for the broad-band application of PCA from a single satellite, greatly increasing the number of potential sources compared to joint observations. Finally, the planned \emph{ATHENA} mission \citep{Barret13} will exponentially increase the number of sources that this method could be applied to, allow studies like this one to be extended to higher redshifts, and enable us to probe spectral variability at much higher frequencies.

\section{Conclusions}
\label{conclusions}

We have analysed a large sample of 26 well studied bright, variable AGN using PCA, isolating the different variable spectral components in a model-independent way. We summarise the key results here:
\begin{itemize}
\item From our sample of 26 different AGN, we find at least 12 qualitatively different patterns of variability on 10~ks time scales. We can match some of these patterns to the predictions from different simple models using simulations, but further work is needed to fully understand all of these results.
\item The variability in almost all sources is dominated a single component, which we find corresponds to the flux of the continuum, generally well modelled by a simple power law.
\item The majority of sources have a second order component that corresponds to a simple spectral pivoting of the primary continuum, without a strong flux dependence. We suggest that this is caused by propagating fluctuations within the corona.
\item We have identified four sources (NGC~4051, NGC~3516, Mrk~766 and 1H~0707-495) which show almost identical variability to that previously investigated in MCG--6-30-15, and demonstrated that this variability can be explained by a relativistic reflection model, ruling out variable absorption as the dominant mechanism of spectral variability.
\item We find three sources (NGC~4395, NGC~1365 and NGC~4151) that show clear evidence of strongly variable partial-covering neutral absorption.
\item Using simulations we have begun to build up a library of PC spectra for different variability mechanisms, including continuum variability under different conditions, neutral and ionised absorption variability, and reflection variability.
\end{itemize}
We conclude that PCA is an extremely powerful and versatile tool for studying the X-ray spectral variability of AGN, and has great potential to contribute to our understanding of these objects, both with current and future missions. We will make our analysis code available upon request.

\section*{Acknowledgements}
MLP acknowledges financial support from the Science and Technology Facilities Council (STFC). This work is based on observations obtained with \emph{XMM-Newton}, an ESA science mission with instruments and contributions directly funded by ESA Member States and the USA (NASA).

\bibliographystyle{mnras}
\bibliography{bibliography_pcapaper}

\clearpage
\appendix

\section{Spectra, Lightcurves and count-count plots}
\label{appendix_spectra}

\begin{figure*}
\centering
\includegraphics[width=16cm]{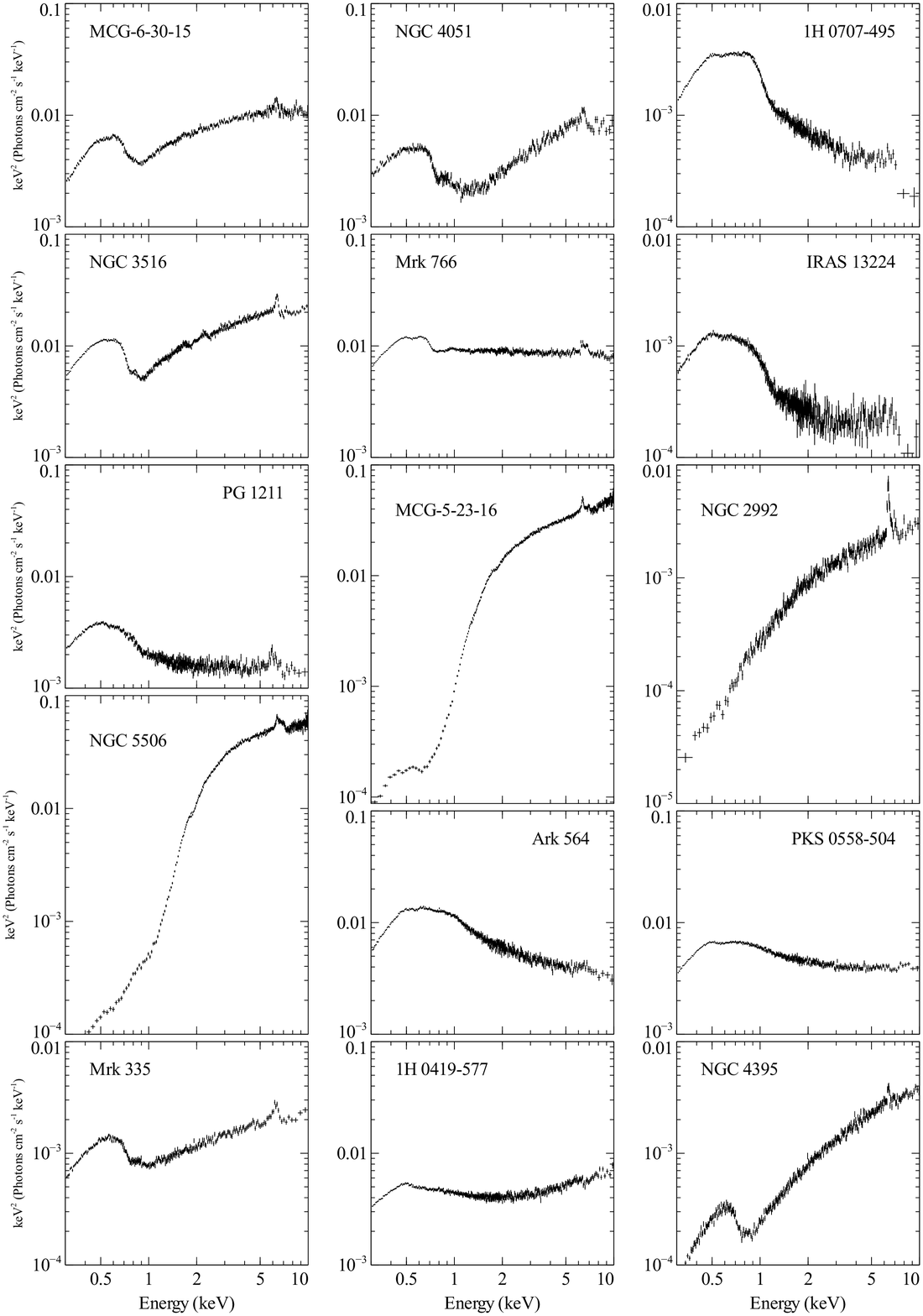}
\caption{Unfolded spectra of the first 15 sources in our sample. Spectra are ordered by the order of sources in the main paper. Some binning is applied in \textsc{Xspec} for clarity. This does not reflect the binning used in the analysis, where all spectra are binned into 50 logarithmic bins for consistency.}
\label{allspectra1}
\end{figure*}

We show here (Figs.~\ref{allspectra1} and~\ref{allspectra2}) representative spectra of the 26 sources in our sample. All spectra are plotted unfolded to a power law with index $\Gamma=0$ and normalization 1. We plot the average spectrum of the longest continuous observation, unless that observation coincides with an atypical flux state. The spectra are not corrected for absorption. This is intended so that the reader can get some idea of the relative data quality in different energy bands.

\begin{figure*}
\centering
\includegraphics[width=16cm]{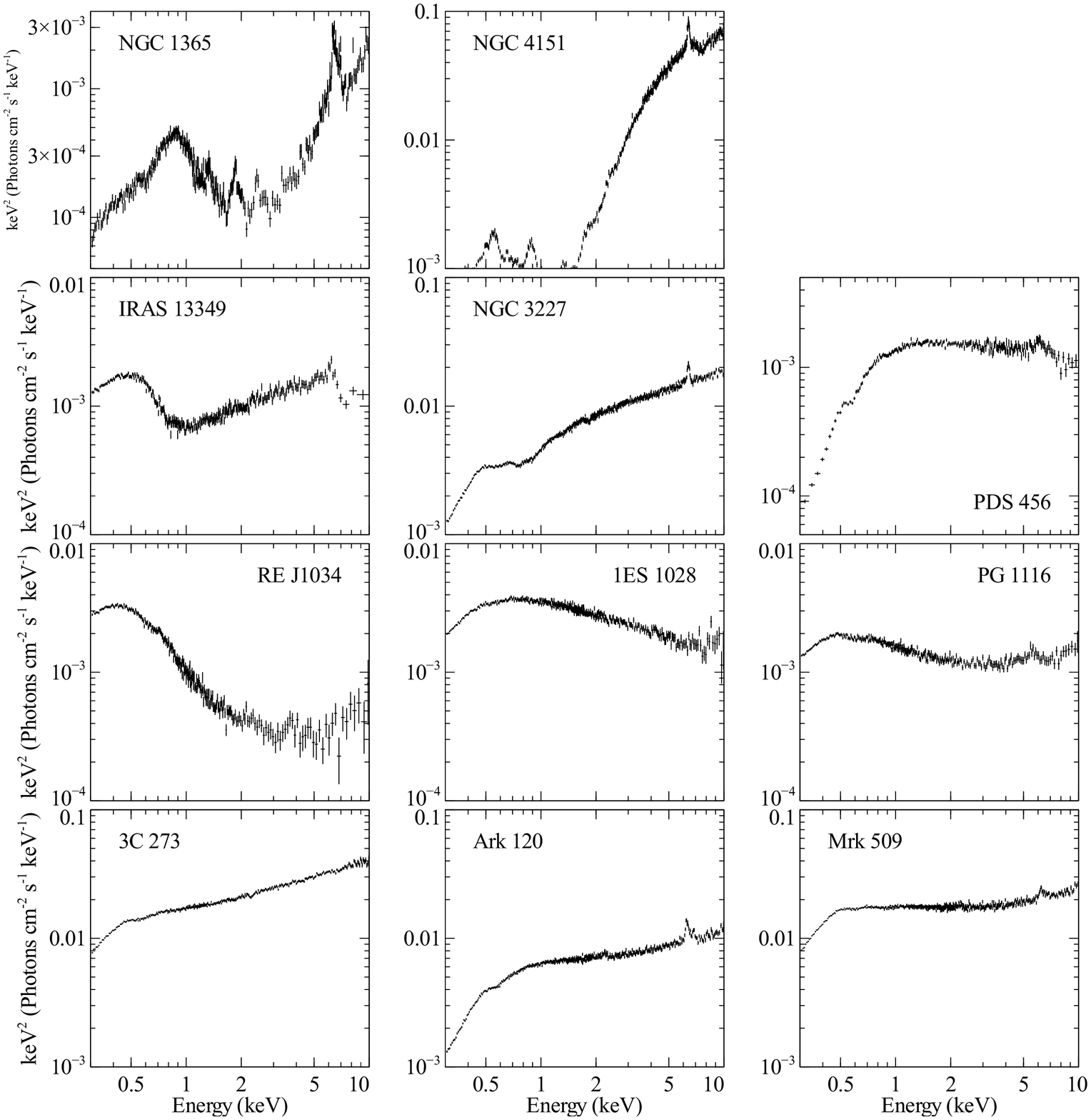}
\caption{Continued from Fig.~\ref{allspectra1}. Unfolded spectra of the remaining 11 sources in our sample.}
\label{allspectra2}
\end{figure*}

\begin{figure}
\centering
\includegraphics[width=8cm]{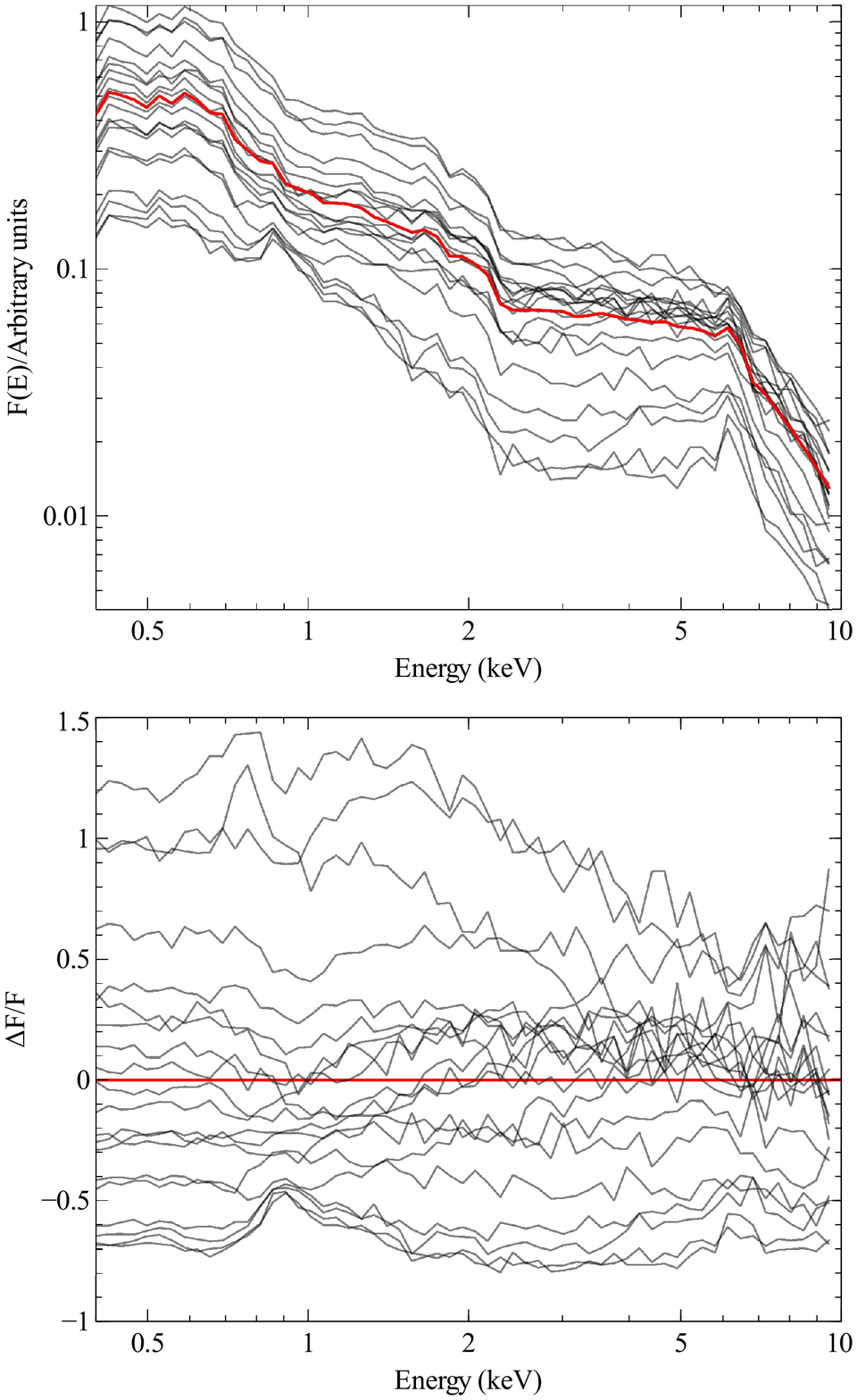}
\caption{Top: A sample of 20 10~ks spectra, used in the analysis of NGC~4051. The grey lines show the individual spectra, and the red line shows the average spectrum of NGC~4051 from all observations. Bottom: The same set of spectra, plotted as fractional residuals to the mean spectrum. This is how the spectra are processed by the PCA code. Error bars are not shown for clarity.}
\label{4051residspectra}
\end{figure}

In Fig.~\ref{4051residspectra} we show a sample of 20 of the 10~ks spectra used in the analysis of NGC~4051. The left panel shows the spectra, with the red line marking the mean spectrum from all observations. The right panel then shows the same set of spectra, plotted as fractional residuals to the mean. The spread in the residual spectra already functions as a crude estimator of the total variability as a function of energy, and a minimum can be seen around the energy of the iron K line.

\begin{figure}
\centering
\includegraphics[width=8cm]{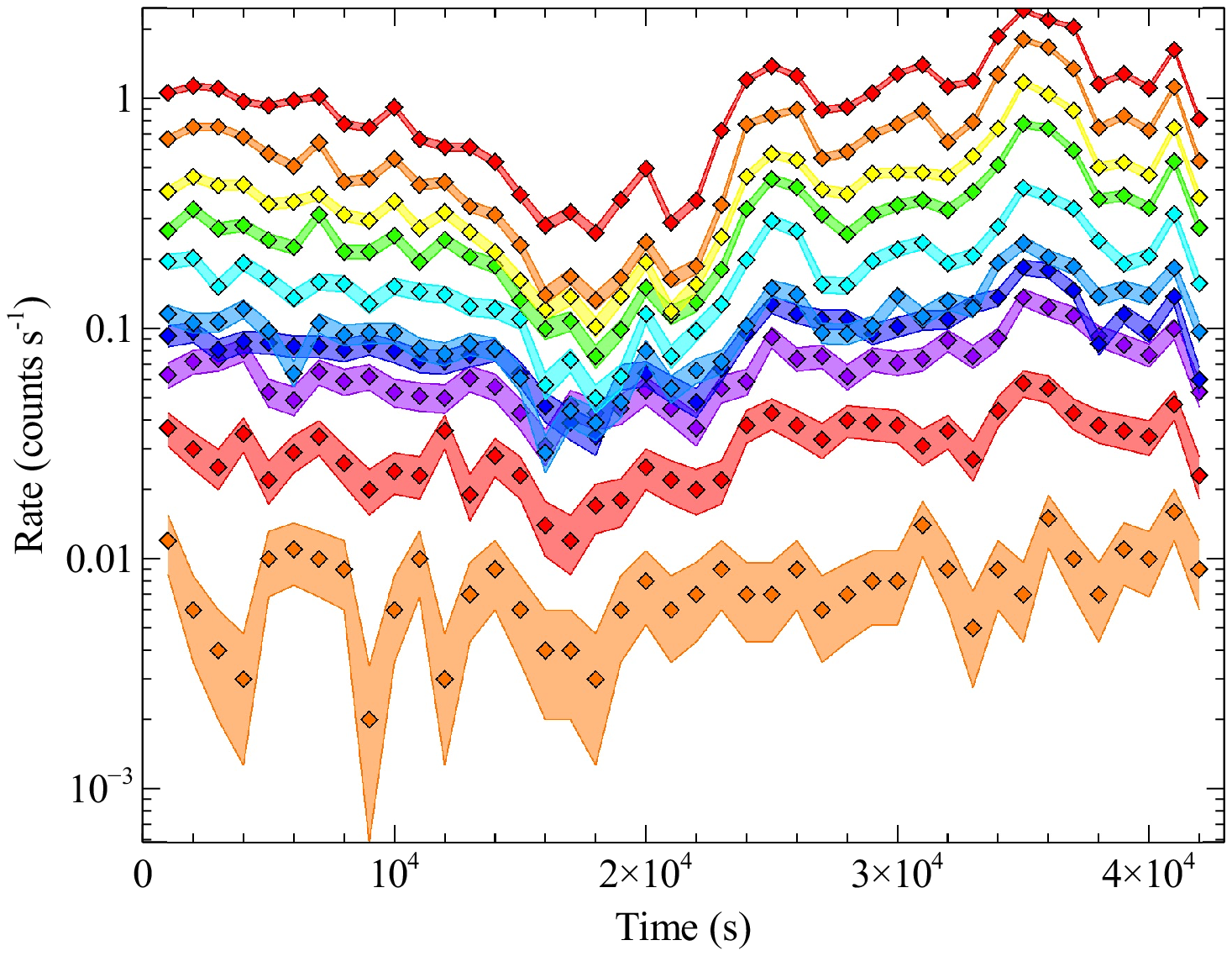}
\caption{Sample of light curves from 10 energy bands from a single observation of NGC~4051. Coloured regions indicate the size of the error bars on each point. Energy increases with decreasing count rate.}
\label{4051lcurves}
\end{figure}

In Fig.~\ref{4051lcurves} we show a sample of 10 light curves from one observation of NGC~4051. For the analysis, we divide the data into 50 logarithmically spaced energy bands, with one spectrum every 10~ks. In the figure, we show every 5th light curve, and increase the sampling to 1~ks.

To give the reader an overview of the amplitude and nature of the variability in each object, we show count-count plots for each source in Figs.~\ref{ccplots1} and~\ref{ccplots2}. These plots simply show the count rate in the 0.5--2~keV band against that in the 2--10~keV band, for each of the 10~ks spectra we use in the analysis. The energies were selected to divide the spectrum approximately equally into soft and hard bands. The points are colour-coded by observation ID. The majority of sources are approximately linear on the count-count plots, with varying degrees of scatter. Two of the objects from group 1 show a down-turn at low fluxes, where the hard flux drops more rapidly than the soft (NGC~4051 and NGC~3516). This has previously been observed in these objects \citep{Taylor03,Noda13} and may be due to spectral pivoting or the variability in different components dominating at different flux levels. The similar downturn in NGC~1365 and NGC~4151 is more likely to be due to the presence of additional, non-nuclear emission at low energies which is constant. Some sources (e.g. 1H~0707-495, PDS~456, PG ~1116, Mrk~509) show several separate linear tracks, corresponding to different observations, which may be due to different components dominating the variability on long and short time-scales, or changes in the absorption of the source spectrum. Generally it seems that the PCs are not strongly affected by differences in the count-count plots between objects within our groups - MCG--6-30-15, NGC~4051 and 1H~0707-495 all have very similar PCs, despite having significant differences in their count-count plots. Similarly, RE~J1034+396, 1ES~1028+511 and PG~1116+215 all have very different count-count plots, but all are divided cleanly into soft and hard components by PCA.

In addition, we show on each count-count plot the orientation of first two eigenvectors, or three if the third is stronger than 1 per cent. These are calculated by multiplying the PCs by the average spectrum for each source, which effectively reinstates the effective area of the detector, then averaging the resultant spectra over the 0.5--2 and 2--10~keV bands. In general, it can be seen that the primary PC is aligned with the bulk of the spectral variability (where it is clearly discernible), and the secondary component describes the scatter about this and is usually close to orthogonal \footnote{We note that the strict orthogonality condition that applies to PCA spectra does not apply here after the renormalization and averaging}. We do not show the third component for most sources as it is an extremely minor effect, and cannot explain the observed scatter in the points. The three sources with extremely hard primary PCs (RE~J1034+396, 1ES~1028+511 and PG~1116+215) stand out when the components are plotted in this way, as unlike all other sources the primary component is aligned almost parallel to the y-axis. 
Another source which stands out is IRAS~13349+2438. Like the other three sources, it has an almost vertical PC1, however this is an effect of the energy band selection, rather than the spectrum being divided into soft and hard components - the first PC in this source is a pivoting term, and crosses the x axis around 1~keV, so the 0.5--2~keV bin averages over positive and negative values, leaving no net flux in this bin.

\begin{figure*}
\centering
\includegraphics[width=16cm]{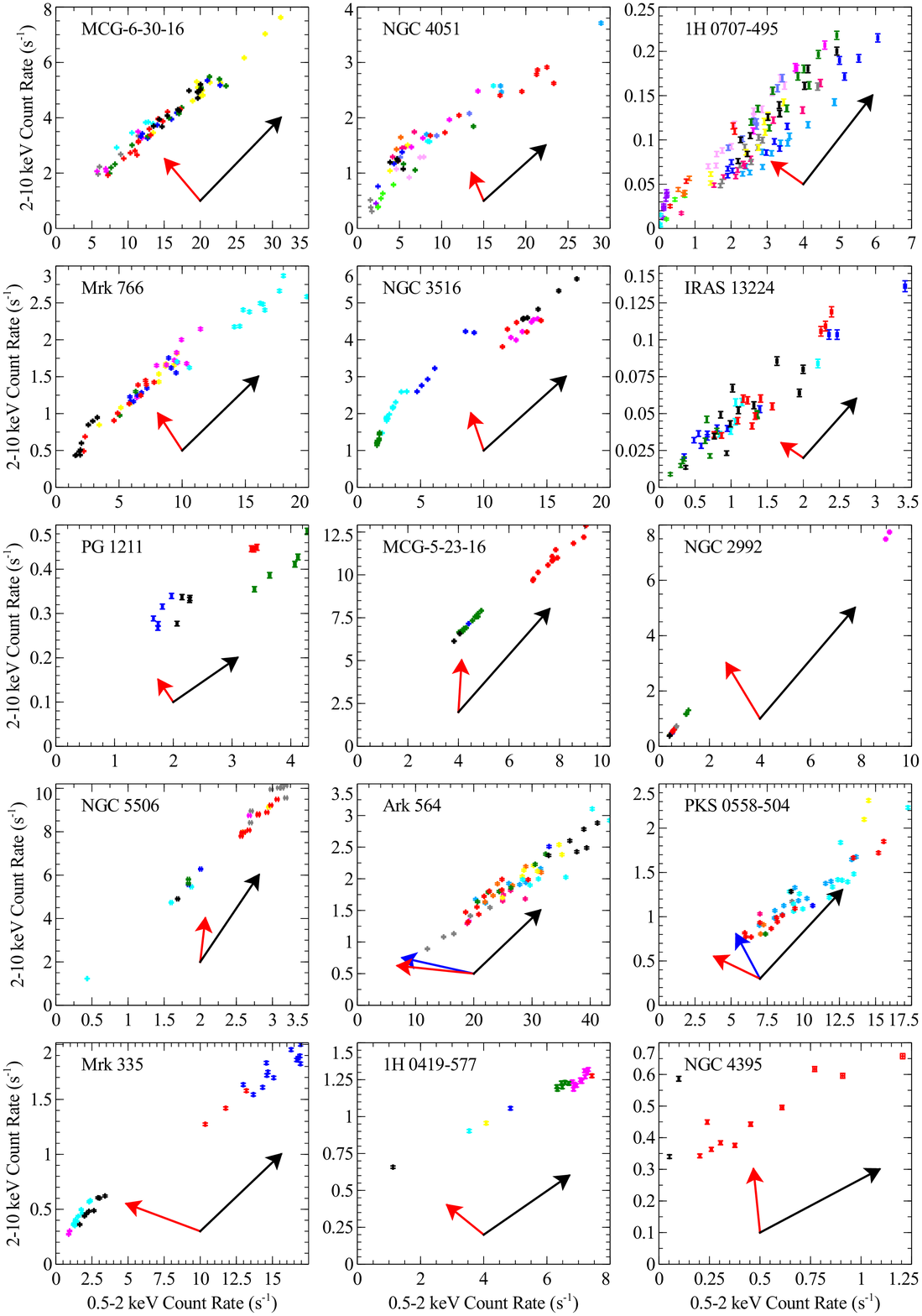}
\caption{Count-count plots for the first 15 sources from our sample. Different observations are shown in different colours. The black and red arrows show the orientation of the first and second PCs, and where the third component is stronger than 1 per cent we show it as a blue arrow. The different sizes of the arrows is not indicative of the strengths of the components, and is intended for clarity only.}
\label{ccplots1}
\end{figure*}

\begin{figure*}
\centering
\includegraphics[width=16cm]{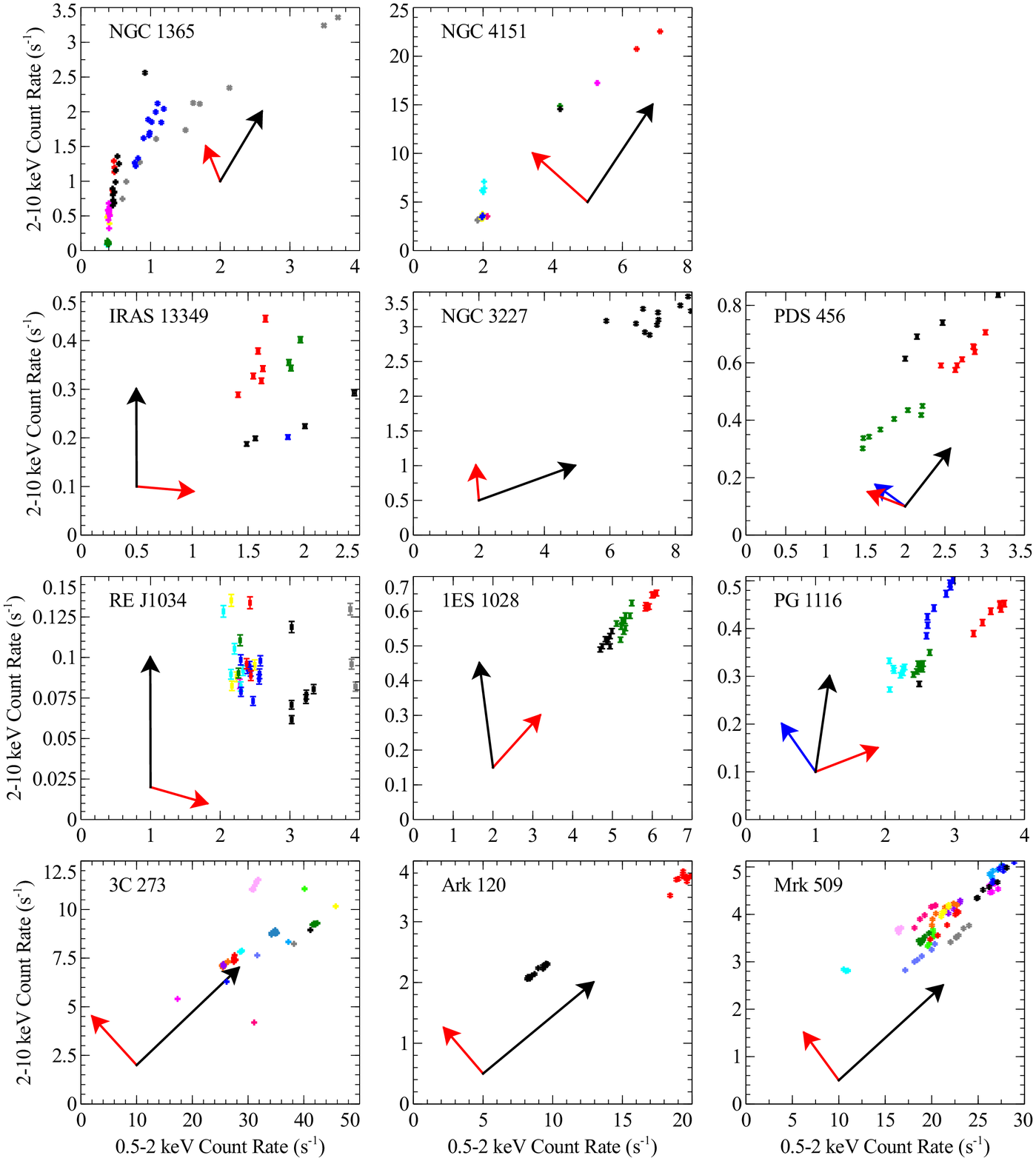}
\caption{Continued from Fig.~\ref{ccplots1}. Count-count plots of the remaining 11 sources in our sample.}
\label{ccplots2}
\end{figure*}

\end{document}